\documentclass[10pt, aps,pra,twocolumn,showpacs,superscriptaddress]{revtex4-2}

\usepackage[colorlinks=true,linkcolor=blue,citecolor=blue,urlcolor=magenta]{hyperref}

\usepackage{amsfonts,mathrsfs,amsmath,amsthm,amssymb}
\usepackage{epsfig}
\usepackage{bm}
\usepackage{array,multirow}
\usepackage{booktabs}
\usepackage{graphicx}
\usepackage{upgreek}

\newcommand{\la}{\langle}
\newcommand{\ra}{\rangle}
\newcommand{\up}{\uparrow}
\newcommand{\dn}{\downarrow}
\newcommand{\om}{\circ}
\newcommand{\xm}{\times}

\begin{document}  
\title {\bf 
Detecting and protecting entanglement  through nonlocality, variational entanglement witness, and nonlocal measurements}

\author{Haruki Matsunaga}
\affiliation{St.Paul's School, London, UK
}

\author{ Le Bin Ho} 
\thanks{Electronic address: binho@fris.tohoku.ac.jp}
\affiliation{Frontier Research Institute 
for Interdisciplinary Sciences, 
Tohoku University, Sendai 980-8578, Japan}
\affiliation{Department of Applied Physics, 
Graduate School of Engineering, 
Tohoku University, 
Sendai 980-8579, Japan}

\date{\today}

\begin{abstract}
We propose an innovative method to enhance the detection and protection of quantum entanglement, a cornerstone of quantum mechanics with applications in computing, communication, and beyond. While entanglement can be represented through nonlocal correlations detectable by the Clauser-Horne-Shimony-Holt (CHSH) inequality, this method does not fully capture all entangled states. To address this limitation, we introduce a variational entanglement witness (VEW) that optimizes the probabilities of detection and improves the efficiency of distinguishing between separable and entangled states. Additionally, we propose a novel nonlocal measurement framework that enables the assessment of both CHSH inequalities and the VEW while preserving the entanglement. Our approach enhances the reliability of entanglement detection while maintaining the entanglement of quantum states, offering significant advancements for quantum technologies.
\end{abstract}
%
%
%\pacs{03.65.Ta, 03.65.Aa, 02.50.-r, 03.67.Ac }
\maketitle

\section{Introduction} 
Entanglement is a fundamental phenomenon where particles share correlated quantum states %that allows particles to be connected 
regardless of their spatial separation \cite{RevModPhys.81.865, 10.1088/978-0-7503-5269-7}. It is crucial for many quantum technologies, such as quantum computing \cite{Yu_2021}, quantum cryptography \cite{Yin2020,doi:10.1126/sciadv.abe6379}, quantum communication \cite{Zou_2021,Wengerowsky2018,doi:10.1126/sciadv.aba4508}, and quantum metrology \cite{10.1063/5.0204102,PhysRevA.94.012339}, among others. However, detecting and protecting entanglement from measurements is significant challenging due to the computationally intractable nature and the fragility of quantum states \cite{GUHNE20091,GURVITS2004448}.

One native approach is full quantum state tomography, 
which provides complete information about the quantum state \cite{PhysRevA.64.052312,PhysRevA.66.012303}. When the quantum state is reconstructed, one can evaluate the entanglement using criteria such as the Peres-Horodecki positive partial transpose (PPT) \cite{PhysRevLett.77.1413,HORODECKI19961} and concurrence \cite{PhysRevLett.78.5022}. However, tomography becomes impractical for large quantum systems due to its exponential scaling with the system size, making it highly resource-intensive.
Recently, Elben et al. introduced a {\it moments of the partially transposed density matrix (PT moments)} protocol, using the first three PT moments to create a simple yet powerful bipartite entanglement test. The measurement was performed using local randomization, eliminating the need for full tomography \cite{PhysRevLett.125.200501}.

An alternative method for detecting entanglement is using entanglement witnesses (EWs), which are operators that identify entanglement by measuring their expectation values \cite{PhysRevLett.92.087902,PhysRevResearch.6.033056,Bae2020,Siudzinska2021,PhysRevApplied.15.054006,Jirakova2024}. EWs offer a way to differentiate between entangled and non-entangled states without quantifying the degree of entanglement. The Bell theorem \cite{PhysicsPhysiqueFizika.1.195} and its associated inequalities, such as the Clauser-Horne-Shimony-Holt (CHSH) inequality \cite{Clauser_1978,PhysRevLett.23.880}, fall within this category. These inequalities detect entanglement by revealing inconsistencies between quantum predictions and local realism. Violations of these inequalities provide evidence of entanglement, making them practical for distinguishing quantum states from classical ones. Although the CHSH inequality is widely used for this purpose \cite{PhysRevA.88.052105,PhysRevA.107.052403,Li2020,Cortes-Vega2023}, it is crucial to note that entanglement can exist even when the inequality is not violated \cite{PhysRevA.40.4277}.

Recent advancements in machine learning have introduced promising methods for detecting quantum entanglement. Neural networks and support vector machines have proven effective in classifying quantum states as either entangled or separable \cite{Urena2024,Asif2023,doi:10.1126/sciadv.add7131,PhysRevApplied.15.054006, Qu2023, PhysRevA.108.052424,ROIK2022128270, Ma2018}, detecting genuine multipartite entanglement \cite{PhysRevA.108.052424}, and developing entanglement witnesses \cite{PhysRevApplied.15.054006,PhysRevResearch.6.033056,PhysRevApplied.19.034058,Ma2018}. So far, convolutional neural networks have been particularly useful for analyzing entanglement patterns \cite{Kookani2024,Qu2023}.
%and classifying entanglement in high-dimensional systems. 
These methods highlight the expanding role of machine learning in enhancing the detection and analysis of quantum entanglement.

% and machine learning \cite{PhysRevApplied.15.054006,PhysRevA.108.052424,Asif2023,Urena2024,Kookani2024, Qu2023,ROIK2022128270,doi:10.1126/sciadv.add7131}.

However, these detection methods rely on multiple measurements of a quantum state and local measurements on spatially separable subsystems, 
which causes the collapse of the global wavefunction of the entire system.
Therefore, there is a need for detection methods that can identify entanglement without destroying it.

In this paper, we %use the CHSH inequality violation as an indicator of entanglement. We also 
introduce a variational entanglement witness (VEW) approach to detect entanglement when the CHSH inequality alone is not enough. While using these quantities for detecting entanglement is widespread, our approach provides an effective tool to confirm a quantum state entangled status. Optimizing VEW also improves the efficiency in distinguishing separable from entangled states.
Moreover, we propose a nonlocal measurement framework to effectively measure the CHSH inequality and VEW, enabling both the detection and protection of entanglement. 

Concretely, in a bipartite system shared with Alice and Bob, we first theoretically examine the entanglement via the violation of the CHSH inequality. We then apply a VEW to independently detect entanglement by training a parameterized witness operator. %regardless of the CHSH criteria. 
%We parameterize the CHSH operator to be the witness operator. 
We extend the application of VEW to general pure and mixed states in two-dimensional systems and further to higher-dimensional systems.
Finally, we propose a nonlocal measurement framework to measure the expectation values in the CHSH and VEW. This framework improves the reliability of entanglement detection and protects it from wave function collapse. We also use superconducting chips to simulate the nonlocal measurement and assess the post-measurement state to confirm the preservation of entanglement. This work not only advances our fundamental understanding of quantum entanglement but also 
supports quantum applications like secure communication and complex computations.

\section{Results} 
\subsection{CHSH inequality and VEW}

\begin{figure}[t]
\includegraphics[width= 8.6cm ]{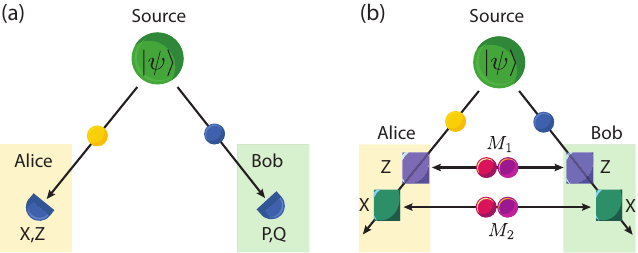}
\caption{
(a) CHSH setup: an entangled bipartite state 
$|\psi\ra$ is divided, with one subsystem 
sent to Alice and the other to Bob. Alice measures 
$X, Z$, while Bob measures 
$P, Q$.
(b) A proposed nonlocal measurement model. 
In this model, two meters M$_1$ and M$_2$ are used to measure 
$ZZ$ and $XX$.}
\label{fig1}
\end{figure}

Suppose Alice and Bob share a bipartite system S represented by $|\psi\ra$ as shown in Fig.~\ref{fig1}(a). Each performs two experiments, with outcomes of either +1 or -1 on their respective parts. Alice measures operators $X$ and $Z$, while Bob measures $P$ and $Q$. The correlation between their measurements is given by the operator
\begin{align}\label{eq:chsh}
    S_{\rm CHSH} = (X + Z)\otimes P + (X - Z)\otimes Q.
\end{align}
In classical scenarios, $X, Z, P, Q \in \{\pm 1\}$ are 
random variables. In each run, $S_{\rm CHSH}$ can be either 
-2 or +2 according to the local hidden variable (LHV) theory. 
The expectation value follows the inequality
\begin{align}\label{eq:vio}
    |\la S_{\rm CHSH}\ra|\le 2.
\end{align}

\begin{figure}[t]
    \includegraphics[width= 8.6cm ]{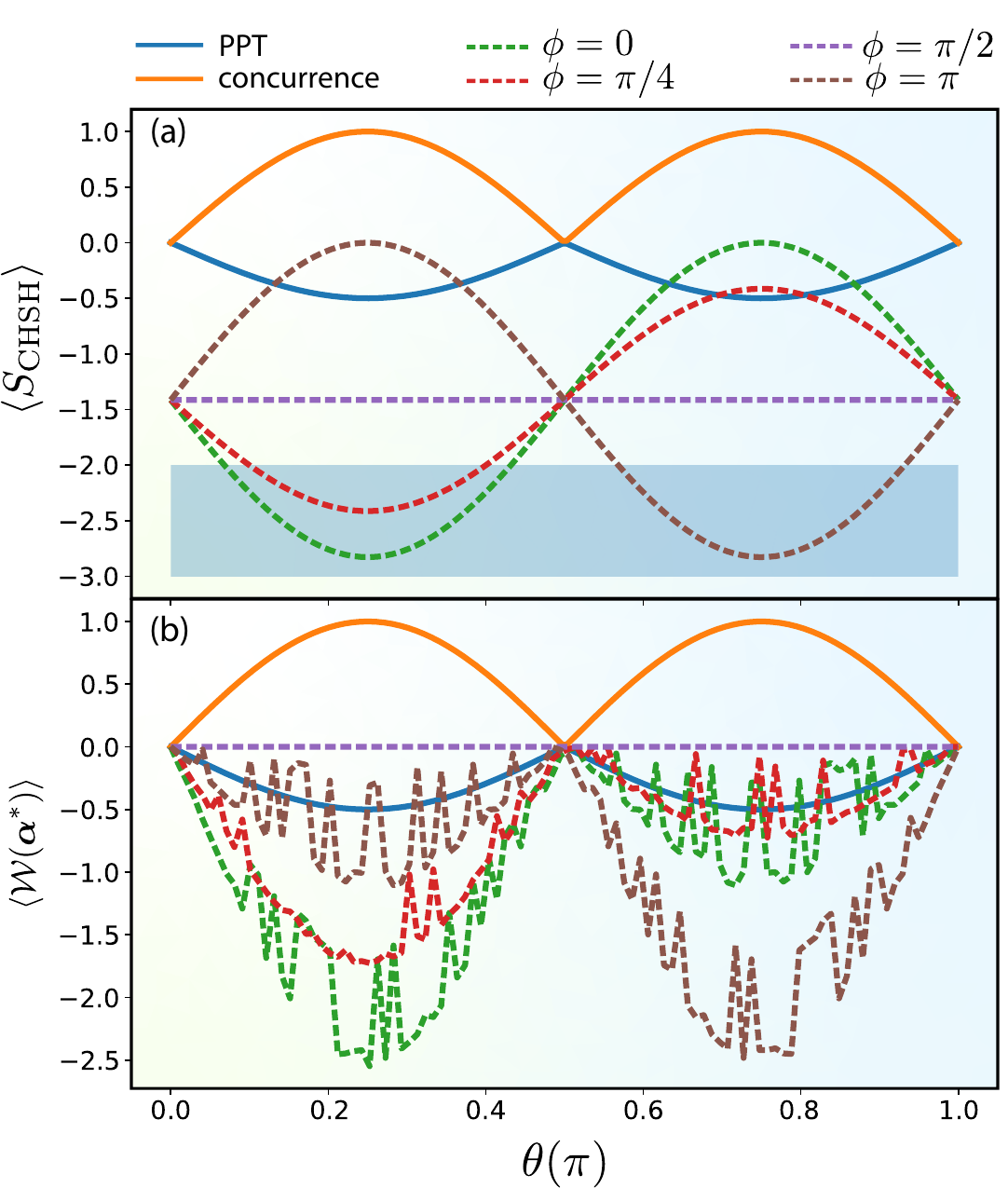}
    \caption{
    (a) Plot of $\la S_{\rm CHSH}\ra$ versus $\theta$ at different $\phi$ values. 
    The CHSH inequality is violated in 
    the aqua region where $|\la S_{\rm CHSH}\ra| > 2$. 
    Entanglement measures including the PPT criterion and 
    concurrence are also shown for comparison.
    (b) Plot of the optimal entanglement witness $\la \mathcal{W} (\bm\upalpha^*)\ra$.}
    \label{fig2}
\end{figure}

In quantum mechanics, 
this inequality can be violated, ie., $|\la S_{\rm CHSH}\ra| > 2$ \cite{Hensen2015,Cirelson1980}.
For example, let Alice and Bob share a quantum state
\begin{align}\label{eq:sys_state}
|\psi\ra = \cos\theta
|00\ra + e^{i\phi} \sin\theta|11\ra,
\end{align}
%where $S$ stands for system, 
where the maximum entanglement occurs at $\theta = \pi/4$.
To maximize the violation of 
the CHSH inequality, 
we choose the Pauli matrices for Alice as
$X = |0\ra\la 1| + |1\ra\la 0|,
Z = |0\ra\la 0| - |1\ra\la 1|$
and for Bob as
$P = -(Z + X)/\sqrt{2}, Q = (Z - X)/\sqrt{2}$ \cite{Cirelson1980}.
A direct calculation of Eq.~\eqref{eq:chsh} yields
\begin{align}\label{eq:}
\notag\la S_{\rm CHSH}\ra %&= \la (X+Z)\otimes P + (X-Z)\otimes Q \ra\\
    &= -\sqrt{2}\big(\la ZZ\ra + \la XX\ra\big) \\
    &= - \sqrt{2}\big(1 + \cos\phi \sin2\theta \big),
\end{align}
where $ZZ$ abbreviates $Z \otimes Z$, and similarly for $XX$.
See the detailed calculation in Appendix~\ref{appA}.
The CHSH inequality \eqref{eq:vio} is violated, ie., $|\la S_{\rm CHSH}\ra| > 2$ at certain $\theta$ and $\phi$,
as shown in Fig.~\ref{fig2}(a). 
The aqua area marks the violation of the CHSH inequality, which also indicates the violation of the LHV model and demonstrates the nonlocality. 

From this nonlocality, we can infer the entanglement in the quantum state \( |\psi\ra \). 
We confirm the entanglement using the PPT criterion and concurrence. Detailed calculations can be found in Appendix \ref{appB}.
As depicted in Fig.~\ref{fig2}(a), the maximum violation of inequality \eqref{eq:vio} corresponds to the highest level of entanglement, indicated by either minimum PPT or maximum concurrence. Thus, the violation of the CHSH inequality indicates the presence of entanglement. 

While the nonlocal behavior can indicate entanglement, however, entanglement does not always imply nonlocality, meaning nonlocality cannot fully detect entanglement.

When the nonlocality fails to detect entanglement, an EW \cite{Chr_2014} is employed. An EW is a Hermitian operator \( \mathcal{W} \) used to assess whether a quantum state \( \rho \) is entangled. If \(\text{Tr}(\mathcal{W} \rho) < 0\), the state is identified as entangled; otherwise, if \(\text{Tr}(\mathcal{W} \rho) \geq 0\), the state is considered non-entangled under this witness. In such a case, a different witness is needed for further evaluation.

An EW may require complete knowledge of the quantum state \cite{RevModPhys.81.865,doi:10.1126/sciadv.add7131}. For example, a general method to construct a witness involves the expression 
\(
\mathcal{W} = O - \min_{\rho \in \{\rho_1\otimes\rho_2\}} {\rm Tr}[O\rho],
\)
where \( O \) is an arbitrary operator, or a witness based on the PPT criterion can be defined by \( \mathcal{W} = (|v\rangle\langle v|)^{T_A} \), where \( |v\rangle \) is the eigenvector of \( \rho^{T_A} \) corresponding to the smallest eigenvalue, and \( T_A \) denotes partial transposition \cite{PhysRevLett.84.2726}.
For pure quantum states, \( |\psi\rangle \), a projector witness for bipartite systems is given by \( \mathcal{W} = \lambda_\psi^2 \bm{I} - |\psi\rangle\langle|\psi| \), where \( \lambda_\psi \) is the largest Schmidt coefficient \cite{PhysRevLett.92.087902}. In the case of two-qubit systems, a correlation-based EW can be expressed as 
\(
\mathcal{W} = \sum_k \sigma_k \otimes \sigma_k,
\)
with \( \sigma_k \) representing the Pauli matrices for \( k \in \{0,x,y,z\} \) \cite{PhysRevA.71.010301,PhysRevA.70.062113}.

%The witness cannot be directly measured locally, it can be approximated using a completely positive map \cite{PhysRevLett.89.127902}, which is equivalent to performing full quantum state tomography \cite{PhysRevA.66.052315,PhysRevLett.116.230403}.

To find the most effective EW, 
here we propose a variational entanglement witness (VEW), 
%for the first time, 
which optimizes \( \mathcal{W} \) 
to better detect entanglement for a given quantum state.
The variational scheme proceeds as follows. 
First, the entanglement witness \( \mathcal{W} \) 
is parameterized by \( \bm{\upalpha} \), ie., \( \mathcal{W}(\bm{\upalpha}) \). 
The expectation value of this witness is then used as the cost function
\begin{align}\label{eq:cost}
     \mathcal{C}(\bm{\upalpha}) = \text{Tr}\big[\mathcal{W}(\bm{\upalpha}) \rho\big].
\end{align}
The optimization process is defined by
\begin{align}
\bm{\upalpha}^* &= \arg\min_{\{\bm{\upalpha}\}} \mathcal{C}(\bm{\upalpha}) \label{eq:opt}\\
\text{s.t.} \quad \mathcal{C}(\bm{\upalpha}^*) &= 0 \quad \forall \text{ separable states } \rho_{\rm sep}. \label{eq:opt1}
\end{align}
The objective is to find \( \bm{\upalpha}^* \) that minimizes the cost function, 
ideally achieving a negative value for entangled states.

In principle, VEW can be constructed from any operator, including Bell's operator, 
however it requires perfect correlations, which are impractical experimentally.
Here, we employ \( S_{\rm CHSH} \) as the variational witness operator, 
thereby removing the requirement for prior knowledge of 
the quantum state and enabling direct measurement. 
The variational witness operator is defined by
\begin{align}\label{eq:vwe}
    \mathcal{W}(\bm{\upalpha}) = -\sqrt{2} \big(\alpha_1 ZZ + \alpha_2 XX\big),
\end{align}
with the cost function is
\begin{align}\label{eq:cost_vew}
    \mathcal{C}(\bm{\upalpha}) = -\sqrt{2} \big(\alpha_1 \la ZZ \ra + \alpha_2 \la XX \ra\big).
\end{align}

The optimization process relies on gradient-free methods using 
the COBYLA optimizer. See detailed calculation in Appendix~\ref{appC}. 

\subsubsection{
Demonstration of VEW for a Bell state.}
We first examine VEW for a Bell state given by 
\( |\psi\ra \) in Eq.~\eqref{eq:sys_state}.
The results, illustrated in Fig.~\ref{fig2}(b), 
effectively demonstrate the entanglement of the quantum state.
At $\theta = 0, \pi/2,$ and $\pi$,
the expectation value $\la\mathcal{W}(\bm\upalpha^*)\ra$ is zero, 
indicating zero in the PPT criterion and concurrence. 
At other points, $\la\mathcal{W}(\bm\upalpha^*)\ra$ is negative, 
indicating the entanglement. Here, we achieve a 
100\% success rate, comparable to the performance of 
the machine learning approach \cite{Ma2018}.
For each point, we optimize VEW 
to obtain its minimum value. 
The minimum values are independent of one another, 
so the curves appear unsmooth.
This observation holds for all $\phi$ except when $\phi = \pi/2$, 
because at this specific point, the expectation value 
$\la XX\ra = 0$ and thus $\mathcal{C}(\bm\upalpha)$ becomes a constant. 
According to the constraint in Eq.~\eqref{eq:opt1}, 
this $\mathcal{C}(\bm\upalpha)$ must vanish.

\subsubsection{
Demonstration of VEW for a general pure state.}
For general cases, we examine a generic bipartite pure state represented by 
\begin{align}\label{eq:rps}
|\psi_{\rm gen}\rangle = a|00\rangle + b|01\rangle + c|10\rangle + d|11\rangle,
\end{align}
where \(a\), \(b\), \(c\), and \(d\) 
are complex coefficients that satisfy 
the normalization condition \(|a|^2 + |b|^2 + |c|^2 + |d|^2 = 1\). 
The concurrence is calculated as
\begin{align}
    C = 2|ad - bc|.
\end{align}

\begin{figure}[t]
    \includegraphics[width= 8.6cm ]{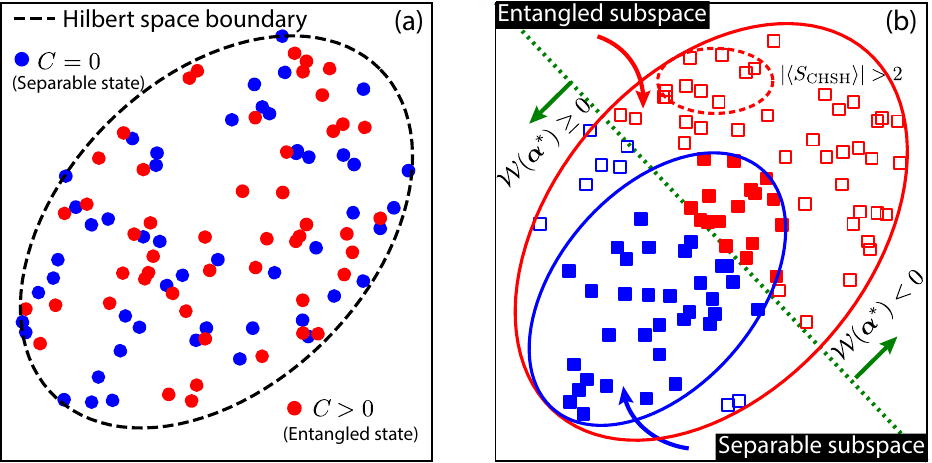}
    \caption{
    (a) A set of 100 random states, 
    with 50 separable states shown as blue dots and 
    50 entangled states as red dots. 
    The states are distributed within 
    an ellipse representing the Hilbert space.  
    (b) Classification results using VEW. 
    The separable subspace is the blue ellipse,
    and the entangled subspace is the region between the red and blue ellipses. 
    Closed squares indicate $C = 0$
    and opened squares are $C > 0$.
    Blue colors indicate \(\langle \mathcal{W} (\bm\upalpha^*) \rangle \ge 0\) (separable), 
    and red colors are \(\langle \mathcal{W} (\bm\upalpha^*) \rangle < 0\) (entangled), 
    separated by the green dotted line \(\langle \mathcal{W} (\bm\upalpha^*) \rangle = 0\).
    %There are four cases: (1) closed blue squares (separable states and have positive VEW), 
    %(2) open red squares (entangled states and have negative VEW), 
    %(3) open blue squares (entangled state but have positive VEW), 
    %and (4) closed red squares (separable states but have negative VEW).
    %Correct classifications are (1, 2), while misclassification occurs 
    %in (3, 4). 
    The inset red dashed ellipse highlights the subspace 
    where the CHSH inequality is violated.
    }
    \label{fig3}
\end{figure}

To evaluate numerically, we generate 100 random states, 
including 50 separable states (\(C = 0\)) 
and 50 entangled states (\(C > 0\)). 
VEW is then applied to classify the states as separable or entangled, 
with the results shown in Fig.~\ref{fig3}.
Refer to App.~\ref{appC} for detailed information on the data generation process.

Figure~\ref{fig3}(a) shows the set of random states, 
with blue dots representing separable states 
and red dots representing entangled states. 
The states are distributed randomly 
within an ellipse that illustrates the Hilbert space.

Figure~\ref{fig3}(b) presents the classification results. 
The separable subspace is a convex subset 
that is nested within a larger convex set of all quantum states, 
while the entangled subspace is the region between them.
Closed squares represent separable states (defined by \(C = 0\)) 
and are placed in the separable subspace. Similarly, 
open squares represent entangled states and are assigned to the entangled subspace.
Using VEW, these states are classified and assigned colors: 
blue means separable states with \(\langle \mathcal{W} (\bm{\upalpha}^*) 
\rangle \geq 0\) and red means entangled states with 
\(\langle \mathcal{W} (\bm{\upalpha}^*) \rangle < 0\). 
The green dotted line indicates \(\langle \mathcal{W} (\bm{\upalpha}^*) 
\rangle = 0\). % and marks the boundary between these classifications.

Correct classifications includes closed blue squares (separable states with positive VEW) and open red squares (entangled states with negative VEW). Misclassifications, on the other hand, appear as open blue squares (entangled states but positive VEW) and closed red squares (separable states but negative VEW). In this example of the random data set, VEW achieves 66\% accuracy for separable states and 84\% for entangled states. Moreover, states violating the CHSH inequality, marked by the smaller red dashed convex subset, are identified as entangled by both VEW and CHSH.

After all, we emphasize that both the CHSH inequality 
and VEW require nonlocal measurements of the expectation values 
\(\la ZZ \ra\) and \(\la XX \ra\). 
We will now present a framework for nonlocal measurements of 
\(\la ZZ \ra\) and \(\la XX \ra\) to confirm the CHSH inequality and VEW.

\subsection{Nonlocal measurement framework for measuring the CHSH inequality and VEW}
To measure $X$ and $Z$ on the Alice's side and $P$ and $Q$ on the Bob's side, 
we need to measure the nonlocal products $Z\otimes Z$ 
and $X \otimes X$, or $ZZ$ and $XX$ for short. 
This is challenging because measuring noncommutative observables 
in a local state is impossible. To overcome this problem, 
we use entangled %qubit pairs as 
meters to couple to both Alice's and Bob's sides and readout the meters' outcomes. 
The initial meters states are maximally entangled Bell states. 
Alice and Bob each couple their subsystem with 
the meters. %qubit using a CNOT (CX) gate. 
After the interactions, they measure 
their meter in the $Z$ bases.

The nonlocal measurement model is shown in Fig.~\ref{fig1}(b). 
To measure $\la ZZ\ra$ and $\la XX\ra$
simultaneously, we use two meters, $\rm M_1$ and $\rm M_2$. 
The meter states are given by Bell states as
\begin{align}\label{eq:meter_state}
|\xi\ra_1 &= \dfrac{1}{\sqrt{2}}\Big(|\up\up\ra + |\dn\dn\ra\Big)\\ %\text {and }
|\xi\ra_2 &= \dfrac{1}{\sqrt{2}}\Big(|\om\!\xm\ra + |\xm\!\om\ra\Big),
\end{align}
where we used $\{|\up\ra,|\dn\ra\}$ 
as the computational basis for M$_1$ 
and $\{|\om\ra,|\xm\ra\}$ for M$_2$, 
which are equivalent to $\{|0\ra,|1\ra\}$ in system S.

\begin{figure*}[t]
    \includegraphics[width= 16cm ]{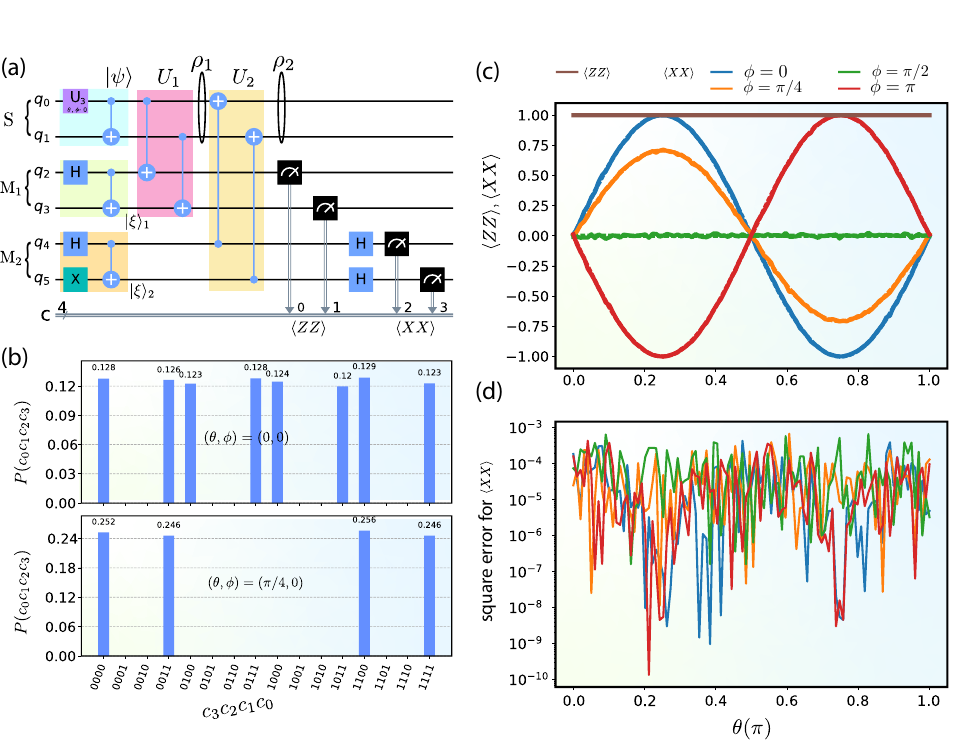}
    \caption{
    (a) Quantum circuit for nonlocal measurement: System S consists of two qubits,
    $q_0$ and $q_1$, meter M$_1$ consists of $q_2$ and $q_3$, 
    and meter M$_2$ consists of $q_4$ and $q_5$.
    (b) Probability $P(c_0c_1c_2c_3)$ plotted for two cases: $(\theta, \phi) = (0,0)$ and $(\pi/4,0)$.
    (c) Plot of $\la ZZ\ra$ and $\la XX\ra$ versus $\theta$ at different $\phi$ values. 
    The dotted curves are theoretical predictions, 
    and the solid curves are from simulations. 
    (d) Corresponding square error between simulations and theory plotted for $\la XX\ra$ case.}
    \label{fig4}
\end{figure*}

%The initial joint system-$M_1$ state is given by
%%\begin{widetext}
%\begin{align}\label{eq:sys_meter_i}
%\notag |\Theta\ra_i &= 
%    |\psi\ra\otimes|\xi\ra_1\\
%    \notag &= \dfrac{1}{\sqrt{2}}\cos\theta |00\ra
%        \Big(|\up\up\ra + |\dn\dn\ra\Big)\\
%        &\hspace{0.5cm}+\dfrac{1}{\sqrt{2}}
%        e^{-i\phi} \sin\theta|11\ra
%        \Big(|\up\up\ra + |\dn\dn\ra\Big).      
%\end{align}
%\end{widetext}
To measure $\la ZZ\ra$ 
(note that hereafter,
we limit ourselves to the expectation values 
w.r.t the quantum state $|\psi\ra$ in Eq.~\eqref{eq:sys_state}), 
we apply the interaction $U_1$ 
between the system and M$_1$ which are two CNOT (CX) gates
as shown in Fig.~\ref{fig4}(a).
%The joint system evolves to
%\begin{align}\label{eq:sys_meter_f}
%|\Theta\ra_1 = 
%   \dfrac{1}{\sqrt{2}}
%   \Big(\cos\theta |00\ra + e^{-i\phi} \sin\theta|11\ra\Big)\otimes
%    \Big(|\up\up\ra + |\dn\dn\ra\Big)
%\end{align}
%
%Perform projection measurement in the meter with the
The measurement observables are represented by Kraus operators as
\begin{align}\label{eq:Mij}
M_{\mu} =\la\mu|U_1|\xi\ra_1,\
\forall \mu \in\{\up\up, \up\dn, \dn\up, \dn\dn\},
\end{align}
which gives 
\begin{align}\label{eq:Mijs}
M_{\up\up} &= M_{\dn\dn} =  
\dfrac{1}{\sqrt{2}}\big(|00\ra\la 00|+|11\ra\la11|\big), \\
M_{\up\dn} &= M_{\dn\up} = 0.
\end{align}
%where $A$ and $B$ stand for Alice and Bob side, respectively. 
Refer to Appendix~\ref{appD} for detailed calculations. 
From these measurements, we obtain the probabilities
$ P_{\mu} = \la\psi|M_{\mu}^\dagger M_{\mu}|\psi\ra$ as
\begin{align}\label{eq:Pijs}
P_{\up\up} = P_{\dn\dn} = \dfrac{1}{2};\ 
P_{\up\dn} = P_{\dn\up} = 0.
\end{align}
Then, the expectation value $\la ZZ\ra$ yields
\begin{align}\label{eq:ZZ}
\la ZZ\ra = 
P_{\up\up}-P_{\up\dn}-P_{\dn\up}
+P_{\dn\dn} = 1.
\end{align}

Even though $\la ZZ\ra = 1$ regardless of $\theta$ and $\phi$
however, this value is not always known beforehand in other scenarios. 
Therefore, using $U_1$ is necessary to transfer information from 
the system to the meter $M_1$, 
enabling the measurement outcomes 
to reveal information about 
the system.

Similarly, we measure $\la XX\ra$ using meter M$_2$, 
which interacts with the system through $U_2$,
represented by two inverted CX gates as shown in Fig.~\ref{fig4}(a). 
The measured operators are given by
\begin{align}\label{eq:No}
   N_{\om\om} &= -N_{\xm\xm} = \dfrac{1}{2\sqrt{2}}
    \big(IX + XI\big),\\
   -N_{\om\xm} &= N_{\xm\om} = \dfrac{1}{2\sqrt{2}}
    \big(IX - XI\big),
\end{align}
and the corresponding probabilities are
\begin{align}\label{eq:pro2}
    P_{\om\om} &= P_{\xm\!\xm} 
    %= \la\psi|E_{\om\om}|\psi\ra
    = \dfrac{1}{4}\big(1+\cos\phi\sin2\theta\big),\\
    P_{\om\xm} &= P_{\xm\om} 
    %= \la\psi|E_{\om\xm}|\psi\ra
    = \dfrac{1}{4}\big(1-\cos\phi\sin2\theta\big).
\end{align}
Finally, the expectation value $\la XX\ra$ is calculated as
\begin{align}\label{eq:XX}
\la XX\ra = 
P_{\om\om}-P_{\om\xm}-P_{\xm\om}
+P_{\xm\xm}
 = \cos\phi \sin2\theta.
\end{align}
Refer to detailed calculations in Append~\ref{appD}.
Consequently, both $\la ZZ \ra$ and $\la XX \ra$ can be measured through nonlocal measurements, thereby verifying %the violation of 
the $S_{\text{CHSH}}$ inequality, %which also implies nonlocality in the system S, 
and can also be used for optimizing the VEW.

\subsection{Simulation on superconducting chip}
We design a quantum circuit as shown 
in Fig.~\ref{fig4}(a) for measuring $\la ZZ\ra$ 
and $\la XX\ra$. System S is $q_0q_1$, 
meter M$_1$ is given by $q_2q_3$, 
and meter M$_2$ is given by $q_4q_5$. 
The system state $|\psi\ra$
is prepared by applying a quantum gate U$_3$ 
onto $q_0$ and a CX gate onto $q_0q_1$, where
\begin{align}\label{eq:U3}
{\rm U}_3(\theta, \phi, \lambda) = \begin{pmatrix}
\cos\frac{\theta}{2} 
& -e^{i\lambda}\sin\frac{\theta}{2} \\
e^{i\phi}\sin\frac{\theta}{2} 
& e^{i(\phi+\lambda)}\cos\frac{\theta}{2}
\end{pmatrix},
\end{align}
where we used $2\theta$ in Eq.~\eqref{eq:U3}
to get $|\psi\ra$.
Similarly, to prepare $|\xi\ra_1$, we apply 
a Hadamard gate onto $q_2$ followed by a CX gate $q_2q_3$, 
and to prepare $|\xi\ra_2$, we apply 
a Hadamard gate onto $q_4$, X gate onto $q_5$, 
followed by a CX gate $q_4q_5$.
The interaction $U_1$ consists of two CX gates, 
while $U_2$ consists of two inverted CX gates 
as shown in the figure.
Measure $q_2, q_3$ in the $Z$ basis gives
the outcome for $\la ZZ\ra$. 
To get $\la XX\ra$, we apply Hadamard gates 
onto $q_4$ and $q_5$, and measure on the $Z$ basis.

For the numerical experiment, we execute the quantum circuit
using Qiskit simulation. 
%The data were taken in June and July 2024. 
For each data point, we run 10000 shots and obtain the 
classical probability $P(c_0c_1c_2c_3)$, which is 
the outcome of the two meters M$_1$ and M$_2$,
where \( c_0c_1 \) are the classical outcomes of M$_1$ and \( c_2c_3 \) are the outcomes of M$_2$.

In Fig.~\ref{fig4}(b), we show $P(c_0c_1c_2c_3)$ 
for several cases of $(\theta, \phi)$, including 
$(\theta, \phi) = (0, 0)$
and $(\theta, \phi) = (\pi/4, 0)$.
First, we emphasize the bases denotation 
in Tab.~\ref{tab1} below.
\begin{table}[h]
\centering
\caption{Bases denotation that used in the meters.}
\begin{tabular}{ @{} c w{c}{1.5cm} w{c}{1.5cm} w{c}{1.5cm} @{} }
\toprule
 & \multicolumn{2}{c@{}}{Bases} \\
\cmidrule(l){2-3} 
  Meter    & 0          & 1 \\
\midrule
M$_1$ & $\up$ & $\dn$\\
\midrule
M$_2$ & $\om$ & $\xm$ \\
\bottomrule
\end{tabular}
\label{tab1}
\end{table}

For example, $c_0c_1c_2c_3 = \text{`}0101\text{'}$ means
$\text{`}\up\dn\om\xm\text{'}$.
With this rule, the probabilities give
\begin{align}\label{eq:pp}
    P_{ij} &= \sum_{\{k,l\} \in \{\om,\xm\}} P(ijkl), \\
     P_{kl} &= \sum_{\{i,j\} \in \{\up,\dn\}} P(ijkl),
\end{align}
where  $\{i,j\} \in \{\up,\dn\}$
and $\{k,l\} \in \{\om,\xm\}$.
Using these probabilities, we can calculate
the nonlocal expectation values 
$\la ZZ\ra$ and $\la XX\ra$.

In Fig.~\ref{fig4}(c), $\la ZZ\ra$ and $\la XX\ra$ 
are shown as functions of $\theta$ for different $\phi$ values. 
Simulation and theoretical results of 
these expectation values are compared and show good agreement. 
The corresponding square errors for $\la XX\ra$  are shown in Fig.~\ref{fig4}(d) which are reasonable. 
These results show the effectiveness of nonlocal measurement in verifying the CHSH inequality and VEW.

\subsection{Post-measurement quantum state}
\begin{figure}[t]
    \includegraphics[width= 8.6cm ]{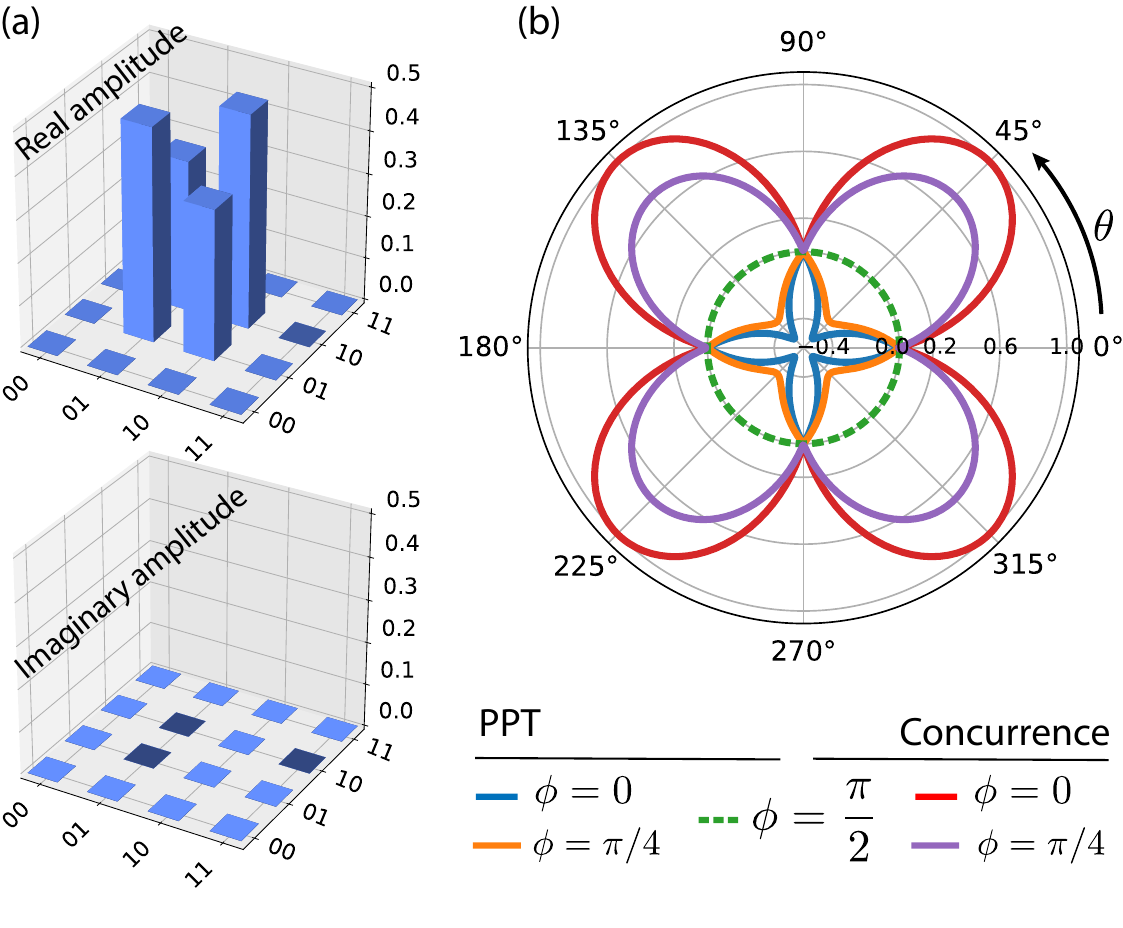}
    \caption{
    (a) Tomography of the final system state at $\theta = \phi = \pi/4$.
    (b) Polar plot of the PPT criterion and concurrence for the final state as functions of $\theta$
    with different $\phi$ values.}
    \label{fig5}
\end{figure}

We derive the system state after these nonlocal measurements.
The system (density) state after 
the first measurement gives \cite{PhysRevA.95.032135}
\begin{align}\label{eq:psi1}
    \rho_1 = \sum_\mu 
    M_\mu|\psi\ra\la\psi|M_\mu^\dagger
    = |\psi\ra\la\psi|,
\end{align}
%or $|\psi_1\ra = |\psi\ra$
see Appendix~\ref{appE} for detailed calculation.
Next, we derive the system state %between the system S and meter M$_2$ 
after the second measurement. It gives
\begin{align}\label{eq:Psi2}
    \rho_2 = \sum_\nu 
    N_\nu\rho_1N_\nu^\dagger
    = \dfrac{1}{2}\begin{pmatrix}
0 & 0 & 0 & 0 \\
0 & 1 & \cos\phi\sin2\theta & 0 \\
0 & \cos\phi\sin2\theta & 1 & 0 \\
0 & 0 & 0 & 0
\end{pmatrix}.
\end{align}
%where $\mu \in 
%\{\om\om, \om\xm, \xm\om, \xm\xm\}$. 
%$N_\mu$ is a measured operator of meter M$_2$,
%and $P_\mu$ is the corresponding probability. 
%
For example, Fig.~\ref{fig5}(a) shows the tomography result 
of the final system state for $\theta = \phi = \pi/4$, 
which matches the theoretical calculation from Eq.~\eqref{eq:Psi2}. 

Finally, we analyze the PPT criterion and concurrence of 
the final state to validate the protection of entanglement. 
For $\theta = k\pi/2$ or  $\phi = \pi/2 + k\pi$
for all $k \in \mathbb{N}$, 
the quantum state is $\rho_2 = {\rm diag}\big(0,1,1,0\big)/2$, 
where both the PPT criterion and concurrence are zero. 
This is depicted by the green dashed line in Fig.~\ref{fig5}(b),
ie., $\phi = \pi/2$. 
The polar plot of PPT criterion and concurrence against 
$\theta$ for different $\phi$ demonstrates 
that maximum entanglement occurs at $\theta = (2k+1)\pi/4$. 
%Remarkably,
%The entanglement %at these points surpasses 
%the original state $|\psi\ra$, 
These results indicate that entanglement is %not only 
preserved %but also strengthened 
under nonlocal measurement.

\subsection{Mixed state case}
In this section, we examine 
$\la S_{\rm CHSH}\ra$ and its indication for
the entanglement in mixed-state cases.
We consider the Werner state as the system state
\begin{align}\label{eq:werner}
\rho = p|\Psi^-\ra\la\Psi^-|+
\dfrac{1-p}{4}\bm{I},
\end{align}
where $0\le p \le 1$ and the Bell states (in the system bases) are defined by
\begin{align}
|\Phi^\pm\ra &= \dfrac{1}{\sqrt{2}}
\Bigl(|00\ra\pm|11\ra\Bigr), \text {and } \label{eq:bell1}\\
|\Psi^\pm\ra &= \dfrac{1}{\sqrt{2}}
\Bigl(|01\ra\pm|10\ra\Bigr). \label{eq:bell2}
\end{align}

Using the same meters M$_1$ and M$_2$ above, 
that means the same POVM as shown in Eqs. (\ref{eq:appEuu}, \ref{eq:appEud}) and Eqs. (\ref{eq:appEoo}, \ref{eq:appEox}).
Then, we have
\begin{align}
    P_{\up\up} &= P_{\dn\dn} = {\rm tr}[\rho E_{\up\up}] = \dfrac{1-p}{4}, \label{eq:mix_Puu}\\
    P_{\up\dn} &= P_{\dn\up} = {\rm tr}[\rho E_{\up\dn}] = \dfrac{1+p}{4}, \and \label{eq:mix_Pud}\\
    P_{\om\om} &= P_{\xm\xm} = {\rm tr}[\rho E_{\om\om}] = \dfrac{1-p}{4},\label{eq:mix_Poo}\\
    P_{\om\xm} &= P_{\xm\om} = {\rm tr}[\rho E_{\om\xm}] = \dfrac{1+p}{4}. \label{eq:mix_Pox}
\end{align}
As a result, we have $\la ZZ \ra + \la XX\ra = -2p$, which implies $\la S_{\rm CHSH}\ra = 2\sqrt{2}p$. The CHSH inequality \eqref{eq:vio} is violated when $1/\sqrt{2} < p \le 1$. As previously stated, this violation region also exhibits entanglement. 

\begin{figure}[t]
    \includegraphics[width= 8.6cm ]{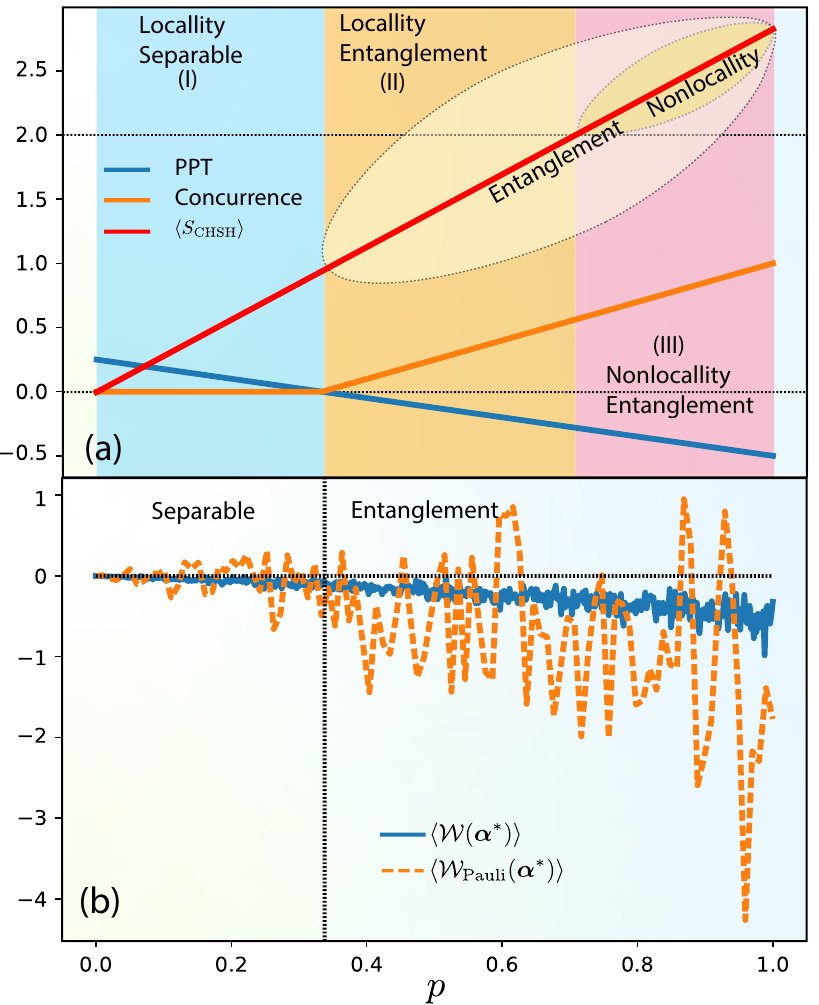}
    \caption{
    (a) Plot of $\la S_{\rm CHSH}\ra$, PPT criterion, and concurrence
    as functions of $p$.
    (b) Plot of the optimal entanglement witness $\la \mathcal{W} (\bm\upalpha^*)\ra$, and $\la \mathcal{W}_{\rm Pauli} (\bm\upalpha^*)\ra$.
    }
    \label{fig6}
\end{figure}

To further explore the entanglement behavior, we again calculate the PPT criterion and concurrence, with results shown in Fig.~\ref{fig6}(a).
First, in region I (sky blue area), $0 \le p \le 1/3$, the system state is local and there is no entanglement, ie., the PPT criterion is positive and the concurrence is zero. In region II (orange area), $1/3 < p \le 1/\sqrt{2}$, the system is local but exhibits entanglement. Finally, in region III (light pink area), $1/\sqrt{2} < p \le 1$, the system state is nonlocal and entangled. This analysis demonstrates that while nonlocal behavior can indicate entanglement, the reverse is not necessarily true. To further clarify, we provide a Venn diagram indicating the relationship between nonlocality and entanglement.

To fully detect entanglement, we use the VEW \(\mathcal{W}(\bm\alpha)\) as defined in Eq.~\eqref{eq:vwe} and compare its performance with the standard Pauli case, \(\mathcal{W}_{\rm Pauli}(\bm\alpha) = \alpha_1 XX + \alpha_2 YY + \alpha_3 ZZ\). The results in Fig.~\ref{fig6}(b) show that the Pauli-based witness often fails to distinguish between entanglement and separability. In contrast, the \(\mathcal{W}(\bm\alpha)\) case successfully detects entanglement for \(p \geq 1/3\), ie.,
\(\mathcal{W}(\bm\alpha^*) < 0\), 
 and identifies separable states for \(p < 1/3\), ie.,
\(\mathcal{W}(\bm\alpha^*) \approx 0\). 
However, the differentiation between entangled and separable states becomes ambiguous around the critical value of \(p = 1/3\), making it challenging to conclusively determine the state.

%but it was unable to detect entanglement in this scenario. 
%Consequently, a more specific witness, such as the projector witness \( \mathcal{W} = \lambda_{\Psi^-}^2 \mathbf{I} - |\Psi^-\rangle\langle\Psi^-| \), would be required to successfully identify entanglement. However, this approach requires prior knowledge of the quantum state, and thus it is not general and not considered in this context.

Next, we derive the post-measurement states, which are given through 
\begin{align}\label{eq:werner1}
    \rho_1 = \sum_\mu 
    M_\mu\rho M_\mu^\dagger
    = \rho,
\end{align}
and
\begin{align}\label{eq:werner3}
    \rho_2 = \sum_\nu 
    N_\nu\rho_1N_\nu^\dagger
    = \begin{pmatrix}
\frac{1 + p}{4} & 0 & 0 & -\frac{p}{2} \\
0 & \frac{1 - p}{4} & 0 & 0 \\
0 & 0 & \frac{1 - p}{4} & 0 \\
-\frac{p}{2} & 0 & 0 & \frac{1 + p}{4}
\end{pmatrix}.
\end{align}
See detailed calculation in Appendix~\ref{appF}.
Finally, we observe that the 
degree of entanglement in this state remains unchanged from the initial state 
$\rho$, indicating that entanglement persists under nonlocal measurement.
%\subsection{Multiparties entanglement detection}
%(\textcolor{red}{move it to appendix)}

We next extend the use of VEW to mixed states beyond the Werner state by considering the Bell-diagonal state, a mixture of the four Bell states  
\begin{align}\label{eq:rhog}
    \rho = \sum_{i=1}^4 \lambda_i |\Lambda_i\rangle\langle\Lambda_i|,
\end{align}
where \(\lambda_i\) are eigenvalues obeying \(\sum \lambda_i = 1\), and \(|\Lambda_i\rangle\) are the Bell states defined in Eqs.~(\ref{eq:bell1}, \ref{eq:bell2}). The entanglement of the state is determined by the largest eigenvalue \(\lambda_i\): the state is separable if \(\mathrm{max}(\lambda_i) \leq 0.5\), and entangled otherwise.

\begin{figure}[t]
    \includegraphics[width= 8.6cm ]{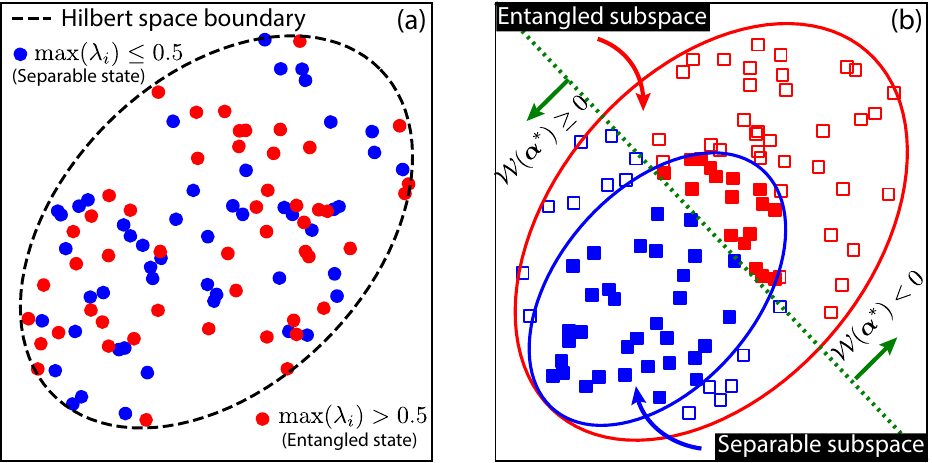}
    \caption{
    Classification of mixed states using VEW, similar to the pure states case in Fig.~\ref{fig3}.  
    (a) States are generated with \({\rm max}(\lambda_i) \le 0.5\) for separable states and \({\rm max}(\lambda_i) > 0.5\) for entangled states.  
    (b) Classification results using VEW.
    }
    \label{fig7}
\end{figure}

Similar to the pure state case in Fig.~\ref{fig3}, we numerically generate 100 random states, evenly split between 50 separable and 50 entangled, by controlling \({\rm max}(\lambda_i)\) as illustrated in Fig.~\ref{fig7}(a). VEW is then applied for classification, with the results shown in Fig.~\ref{fig7}(b). VEW achieves classification accuracies of 66\% for separable states and 70\% for entangled states.
In this example dataset, none of the entangled states violate the CHSH inequality. See also Fig.~\ref{fig9}(b).

\subsection{Demonstration of VEW for high-dimension systems}
We have discussed various applications of the proposed VEW for two-dimensional bipartite systems with pure and mixed states. Now, we apply it to high-dimensional bipartite systems.

We consider a bipartite system of two qudit subsystems in the Hilbert space \( H = \mathbb{C}_A^d \otimes \mathbb{C}_B^d \), with canonical bases \( \{|i,i\rangle \equiv |i_A\ra\otimes|i_B\ra\}_{i=0}^{d-1} \), where \( d \) is the dimension of each qudit. The Bell inequality generalizes to the Collins-Gisin-Linden-Massar-Popescu (CGLMP) inequalities for higher-dimensional systems \cite{PhysRevLett.88.040404}.
We define the CGLMP operator \( S_{\rm CGLMP} \) as
\begin{align}\label{eq:cglmp}
    S_{\rm CGLMP} = (X_d + Z_d)\otimes P_d + (X_d - Z_d)\otimes Q_d,
\end{align}
where the shift and phase operators $X_d$ and $Z_d$ are 
\begin{align}
    X_{kl} =
\begin{cases}
1 & \text{if } l = (k + 1) \mod d, \\
0 & \text{otherwise},
\end{cases}
\end{align}
\begin{align}
    Z_{kl} =
\begin{cases}
\exp\left(\frac{2 \pi i}{d} \cdot k\right) & \text{if } k = l, \\
0 & \text{otherwise},
\end{cases}
\end{align}
and 
$P_d = -(Z_d + X_d)/\sqrt{2}, Q_d = (Z_d - X_d)/\sqrt{2}$.

The common used EW \( \mathcal{W}_d \) is derived from \( S_{\rm CGLMP} \) as
\begin{align}
    \mathcal{W}_d = 2\textbf{I}_d - S_{\rm CGLMP},
\end{align}
where \( \textbf{I}_d\) is the identity operator in \( d \)-dimensional space.
We also define the VEW as
\begin{align}
    \mathcal{W}_d(\bm{\upalpha}) = -\sqrt{2} \left( \upalpha_1 Z_d \otimes Z_d + \upalpha_2 X_d \otimes X_d \right),
\end{align}
and the cost function
$\mathcal{C}_d(\bm{\upalpha}) = \la \mathcal{W}_d(\bm{\upalpha})\ra$.
The optimization problem is given by
\begin{align}
\bm{\upalpha}^* &= \arg\min_{\{\bm{\upalpha}\}} \mathcal{C}(\bm{\upalpha}) \label{eq:optd}\\
\notag \text{s.t.} \quad \mathcal{C}(\bm{\upalpha}^*) &= 0 \quad \forall \text{ separable states } |i,i\ra
\ \forall i \in \{0, d-1\}.
\end{align}

\begin{figure}[t]
    \includegraphics[width= 8.6cm ]{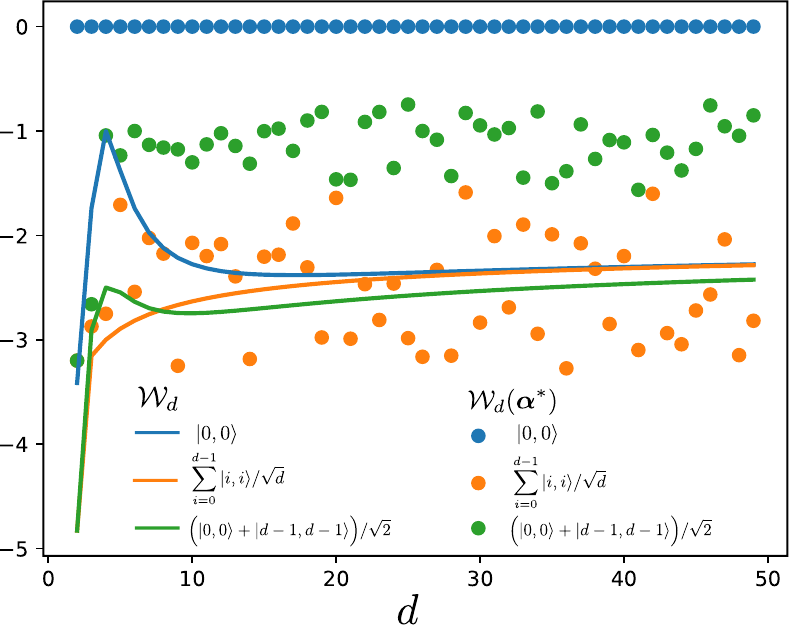}
    \caption{
    Classification of quantum states in high-dimensional systems using EW \( \mathcal{W}_d \) (solid curves) and VEW \( \mathcal{W}_d(\bm{\upalpha}^*) \) (dotted curves): EW fails to distinguish between entangled and separable states, whereas VEW successfully classifies them.
    }
    \label{fig8}
\end{figure}

Figure~\ref{fig8} shows the numerical results for several quantum states, including the separable state \( |\psi\rangle = |0,0\rangle \) and two entangled states: \( |\psi\rangle = \sum_i |i,i\rangle / \sqrt{d} \) (maximum entangled state) and \( |\psi\rangle = (|0,0\rangle + |d-1,d-1\rangle) / \sqrt{2} \) (partial entangled state). We compare the performance of EW \( \mathcal{W}_d \) (solid curves) and VEW \( \mathcal{W}_d(\bm{\upalpha}^*) \) (dotted curves). Although the EW does not distinguish between entangled and separable states (all \( \mathcal{W}_d \) are negative), the VEW successfully classifies them (\( \mathcal{W}_d(\bm{\upalpha}^*) = 0 \) for separable state and
\( \mathcal{W}_d(\bm{\upalpha}^*) < 0 \) for entangled states), demonstrating its advantage in detecting entanglement in high-dimensional systems.

\section{Conclusion} 
We made significant progress in detecting and protecting quantum entanglement using nonlocality, variational entanglement witness (VEW), and nonlocal measurements. While traditional methods like violations of the CHSH inequality are effective, they do not cover all scenarios of entanglement detection. By introducing VEW, we addressed the limitations of these traditional methods, and offered a more comprehensive approach to identifying entanglement. 
%\textcolor{red}{While detecting entanglement using these quantities is widespread, our approach enables confirmation of a quantum state entangled status effectively.}
We also proposed a nonlocal measurement framework for measuring the CHSH operator and VEW.
Our findings emphasize the crucial role of nonlocal measurements in detecting and maintaining entanglement,
which are essential for the functionality of quantum technologies. 
This work not only advances the understanding of quantum entanglement but also contributes to the practical development of more robust quantum computing, communication, and sensing systems. 
%By enhancing the detection and protection mechanisms of entanglement, our research paves the way for more reliable and scalable quantum technologies.

\begin{acknowledgments}
This paper is supported by  
JSPS KAKENHI Grant Number 23K13025.
The code is available at: \href{https://github.com/echkon/nonlocalMeasurement}{https://github.com/echkon/nonlocalMeasurement}
\end{acknowledgments}
%{\it Acknowledgments.} --- 

%\section*{Code availability}

%\begin{widetext}
\appendix

\setcounter{equation}{0}
\renewcommand{\theequation}{A.\arabic{equation}}
\section{CHSH inequality}
\label{appA}

In this Appendix, we derive details of the CHSH inequality.
We start from Eq.~\eqref{eq:chsh} in the main text as
\begin{align}\label{app:chsh}
    S_{\rm CHSH} = (X + Z)\otimes P + (X - Z)\otimes Q.
\end{align}
Using $P = -(Z + X)/\sqrt{2},\ Q = (Z - X)/\sqrt{2}$, we have
\begin{align}\label{app:chsh:1}
    \notag S_{\rm CHSH} &= 
    -(X + Z)\otimes \dfrac{(Z + X)}{\sqrt{2}} 
    + (X - Z)\otimes\dfrac{(Z - X)}{\sqrt{2}}\\
    &=-\sqrt{2}(XX + ZZ).
\end{align}
Now, the expectation value gives
\begin{align}\label{app:chsh:2}
\notag \la S_{\rm CHSH}\ra 
%&= -\sqrt{2}(\la XX  + ZZ\ra)\\ \notag  
&= -\sqrt{2} \la\psi| \big(XX  + ZZ\big)|\psi\ra\\ 
\notag &= -\sqrt{2}\big(\la00|\cos\theta + e^{-i\phi} \sin\theta\la11|\big)\\
\notag &\hspace{0.0cm} \times \Big[(|00\ra+|11\ra)(\la00|+\la 11|) - \\\notag & \hspace{0.5cm} (|01\ra-|10\ra)(\la 01| - \la 10|)\Big] \\
\notag &\hspace{1cm} \times 
\big(\cos \theta|00\ra + e^{i\phi} \sin\theta |11\ra\big)\\
\notag
%&= -\sqrt{2}(\cos^2 \theta + \sin^2 \theta + \dfrac{1}{2}(e^{i\phi}\sin2\theta + e^{-i\phi}\sin2\theta))\\
\notag &= -\sqrt{2}(\cos^2 \theta + \sin^2 \theta + \dfrac{1}{2}(2\cos\phi)\sin2\theta) \\
&= -\sqrt{2}(1 + \cos\phi \sin 2\theta)
\end{align}

\setcounter{equation}{0}
\renewcommand{\theequation}{B.\arabic{equation}}
\section{Entanglement measures}
\label{appB}
We discuss various entanglement measures here, 
including the PPT criterion \cite{PhysRevLett.77.1413, HORODECKI19961} and concurrence \cite{PhysRevLett.78.5022} for bipartite systems. 
Let $\rho$ be a density matrix of an arbitrary mixed state of the two-qubit system AB.
Its partial transpose (with respect to the B party) is defined as
\begin{align}\label{eq:ptB}
    \rho^{T_B} = (I\otimes T)[\rho].
\end{align}
If $\rho^{T_B}$ has a negative eigenvalue, $\rho$ is guaranteed to be entangled. 

For concurrent, we first derive 
$\rho_C$ as
\begin{align}
    \rho_C = (Y\otimes Y)\rho^*(Y\otimes Y),
\end{align}
where $\rho^*$ is the complex conjugate
of $\rho$.
Let $\lambda_1 \ge \lambda_2 \ge \lambda_3 \ge \lambda_4$ are eigenvalues of $\rho_C$,
the concurrence is defined by
\begin{align}
    \mathcal{C}(\rho) 
    = \max \big\{0, \lambda_1 - \lambda_2 - \lambda_3 - \lambda_4\big\}.
\end{align}
The bipartite system is entangled if $\mathcal{C}(\rho) > 0$
and the maximum of $\mathcal{C}(\rho)$ means the maximum of entanglement.

\setcounter{equation}{0}
\renewcommand{\theequation}{C.\arabic{equation}}
\section{Variation entanglement witness}\label{appC}
In this section, we outline the optimization process for the variational entanglement witness. The cost function is defined as
\begin{align}\label{eq:costapp}
    \mathcal{C}(\bm{\upalpha}) = \text{Tr}\big[\mathcal{W}(\bm{\upalpha}) \rho\big],
\end{align}
where \(\mathcal{W}(\bm{\upalpha})\) is a Hermitian operator parameterized by \(\bm{\upalpha}\), and \(\rho\) is the quantum state of interest.

To minimize the cost function, we use the gradient-free COBYLA optimizer. This optimization yields the final parameters \(\bm{\upalpha}^*\), which minimize the cost function \(\mathcal{C}(\bm{\upalpha})\) and represent the optimal parameters for minimizing \(\text{Tr}[\mathcal{W}(\bm{\upalpha}) \rho]\).

For pure states, where \(\mathcal{W}(\bm{\upalpha}) = -\sqrt{2} (\alpha_1 ZZ + \alpha_2 XX)\) and \(\rho = |\psi\ra\la\psi|\), we use the set of separable states \(\{ |00\ra\la00|, |01\ra\la01|, |10\ra\la10|, |11\ra\la11|\}\). The initial parameters are \(\bm{\upalpha} = [4.0, 0.0]\).

For mixed states, we use both the witness operator \(\mathcal{W}(\bm{\upalpha})\) and a Pauli-based witness operator \(\mathcal{W}_{\rm Pauli}(\bm{\upalpha}) = \alpha_1 XX + \alpha_2 YY + \alpha_3 ZZ\). The quantum state \(\rho\) is the Werner state, as defined in Eq.~\eqref{eq:werner}. In addition to the previous separable states, we include \(\rho(p = 0) = \bm{I}/4\). The initial parameters for the Pauli-based case are chosen randomly.

\begin{figure}[h]
    \includegraphics[width= 8.6cm ]{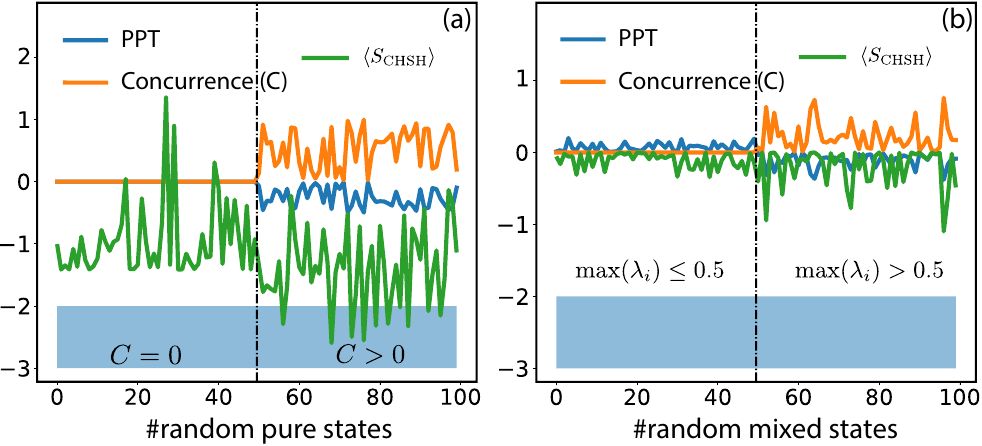}
    \caption{
    (a) Generating of 100 random pure states, with 50 states having \( C = 0 \) 
    and 50 states having \( C > 0 \). For each state, the PPT, concurrence, 
    and \( \langle S_{\rm CHSH} \rangle \) values are displayed. 
    (b) A similar process was applied to mixed states, 
    with 50 states having \( {\rm max}(\lambda_i) \le 0.5 \) 
    and 50 states having \( {\rm max}(\lambda_i) > 0.5 \) .  
    These data are used to examine VEW, 
    as shown in Figs.~(\ref{fig3}, \ref{fig7}) of the main text.
    }
    \label{fig9}
\end{figure}

\textbf{Additional data for general random pure and mixed states.}
For general pure states, 
100 random states were generated according to
Eq.~\eqref{eq:rps}, consisting of 50 separable states ($C = 0$) 
and 50 entangled states ($C>0$). 
The PPT, concurrence, 
and \( \langle S_{\rm CHSH} \rangle \) values were analyzed, 
as shown in Fig.~\ref{fig9}(a).  
Similarly, Fig.~\ref{fig9}(b) presents results for 100 random mixed states, 
generated from Eq.~\eqref{eq:rhog}, with 50 states satisfying 
\( \text{max}(\lambda_i) \leq 0.5 \) and 
50 states satisfying \( \text{max}(\lambda_i) > 0.5 \).  
These data are used to examine VEW, as detailed in Figs.~\ref{fig3} and \ref{fig7} of the main text.

\setcounter{equation}{0}
\renewcommand{\theequation}{D.\arabic{equation}}
\section{Nonlocal measurement}
\label{appD}
In this section, we derive detailed calculations of nonlocal measurement of $\la ZZ\ra$ and $\la XX\ra$.
\subsection{Nonlocal measurement of \texorpdfstring{$\la ZZ\ra$}{ }}
The measurement of $\la ZZ\ra$ is given through meter
M$_1$ is initially prepared in 
$|\xi\ra_1 = \dfrac{1}{\sqrt{2}}\big(|\up\up\ra + |\dn\dn\ra\big)$. 
First, the interaction between the system 
and the meter M$_1$ is given through $U_1$, which
are CX gates as
\begin{widetext}
\begin{align}
\notag  {\rm CX}_{q_0q_2}\ {\rm CX}_{q_1q_3}
            & = \Big(|0\ra\la0|_{q_0}\otimes I_{q_2} 
                + |1\ra\la1|_{q_0}\otimes X_{q_2}\Big)
                \Big(|0\ra\la0|_{q_1}\otimes I_{q_3} 
                + |1\ra\la1|_{q_1}\otimes X_{q_3}\Big) \\
& = \Big(|00\ra\la00|_{q_0q_1}\otimes I_{q_2}I_{q_3} 
            + |01\ra\la01|_{q_0q_1}\otimes I_{q_2}X_{q_3} 
            + |10\ra\la10|_{q_0q_1}\otimes X_{q_2}I_{q_3} 
            + |11\ra\la11|_{q_0q_1}\otimes X_{q_2}X_{q_3}\Big),
\end{align}
%\end{widetext}
where CX$_{q_iq_j}$ is a CNOT gate
with the control qubit is $q_i$ 
and the target qubit is $q_j$. 
The action of $U_1$ onto $I_{q_0}\otimes I_{q_1}\otimes|\xi\ra_1$ gives
(hereafter, we omit $I_{q_0}\otimes I_{q_1}$ and the subscripts $q_i$ for short)
%\begin{widetext}
\begin{align}\label{eq:app_U1xi1}
\notag U_1|\xi\ra_1 &=
    \Big(
    |00\ra\la00|\otimes II + |01\ra\la01|\otimes IX + |10\ra\la10|\otimes XI + |11\ra\la11|\otimes XX\Big)\otimes\dfrac{1}{\sqrt{2}}\big(|\up\up\ra + |\dn\dn\ra\big)\big) \\
    &=\dfrac{1}{\sqrt{2}}
    \Big(|00\ra\la00|+|11\ra\la11|\Big)\otimes
    \Big(|\up\up\ra+|\dn\dn\ra\Big)
    +
    \Big(|01\ra\la01|+|10\ra\la10|\Big)\otimes
    \Big(|\up\dn\ra+|\dn\up\ra\Big).
\end{align}    
\end{widetext}

The measure observables are given in
Kraus operators as
$M_{\mu} = \la \mu|U_1|\xi\ra_1, 
\forall \mu \in\{\up\up, 
\up\dn, \dn\up, \dn\dn\}$, where
\begin{align}
 M_{\up\up} &= \la \up\up| U_1|\xi\ra_1
= \dfrac{1}{\sqrt{2}}\big(|00\ra\la 00|+|11\ra\la11|\big), \label{eq:appM_uu} \\
M_{\up\dn} &= \la \up\dn| U_1|\xi\ra_1
= \dfrac{1}{\sqrt{2}}\big(|01\ra\la 01|+|10\ra\la10|\big), \label{eq:appM_ud} \\
M_{\dn\up} &= \la \dn\up| U_1|\xi\ra_1
= \dfrac{1}{\sqrt{2}}\big(|01\ra\la 01|+|10\ra\la10|\big), \label{eq:appM_du} \\
M_{\dn\dn} &= \la \dn\dn| U_1|\xi\ra_1
= \dfrac{1}{\sqrt{2}}\big(|00\ra\la 00|+|11\ra\la11|\big). \label{eq:appM_dd} 
\end{align}
We next calculate the positive operator-valued measure (POVM) $E_\mu = M_{\mu}^\dagger M_{\mu}$, which give 
\begin{align}
    E_{\up\up} = E_{\dn\dn} = M_{\up\up}^\dagger M_{\up\up} = \dfrac{1}{2}\big(|00\ra\la 00|+|11\ra\la11|\big), \label{eq:appEuu} \\
    E_{\up\dn} = E_{\dn\up} = M_{\up\dn}^\dagger M_{\up\dn} = \dfrac{1}{2}\big(|01\ra\la 01|+|10\ra\la10|\big). \label{eq:appEud}
\end{align}
And the probabilities yield 
\begin{align}\label{eq:app_P_uu}
\notag P_{\up\up} &= \la\psi|E_{\up\up}|\psi\ra \\
\notag &= \big(
        \cos\theta\la00|+e^{-i\phi}\la11|
        \big)
%\notag  &\hspace{0.5cm}\times 
        \dfrac{1}{2}\big(|00\ra\la 00|+|11\ra\la11|\big)\\
\notag  &\hspace{3.0cm} \times\big(
        \cos\theta|00\ra+e^{i\phi}|11\ra
        \big)\\
        &= \dfrac{1}{2},
\end{align}
and 
\begin{align}\label{ZZ}
P_{\up\dn} &= P_{\dn\up} = \la\psi|E_{\up\dn}|\psi\ra = 0.\\
P_{\dn\dn} &= \la\psi|E_{\dn\dn}|\psi\ra = \dfrac{1}{2}.
\end{align}
Finally, we obtain the expectation value
\begin{align}\label{eq:appZZ}
\la ZZ\ra = 
P_{\up\up}-P_{\up\dn}-P_{\dn\up}
+P_{\dn\dn} = 1,
\end{align}
as shown in Eq.~\eqref{eq:ZZ} in the main text.

\subsection{Nonlocal measurement of \texorpdfstring{$\la XX\ra$}{ }}
Similar, to measure $\la XX\ra$,
we use meter M$_2$ with the initial state
$|\xi\ra_2 = \dfrac{1}{\sqrt{2}}
\big(|\om\!\xm\ra + |\xm\!\om\ra\big)$.
The interaction $U_2$ is the inverted CX gates
\begin{widetext}
\begin{align}\label{eq:appinverCNOT}
  \notag \bar{\rm CX}_{q_0q_4} \bar{\rm CX}_{q_1q_5}
  &=\big(I_{q_0}\otimes|\om\ra\la\om|_{q_4}
        + X_{q_0}\otimes|\xm\ra\la\xm|_{q_4}\big)
  \big(I_{q_1}\otimes|\om\ra\la\om|_{q_5}
        + X_{q_1}\otimes|\xm\ra\la\xm|_{q_5}\big)\\
   &= II \otimes |\om\!\om\ra\la\om\!\om| + 
   IX \otimes |\om\!\xm\ra\la\om\!\xm| +
   XI \otimes |\xm\!\om\ra\la\xm\!\om| + 
   XX \otimes |\xm\!\xm\ra\la\xm\!\xm| ,
\end{align}
\end{widetext}
where $q_4$ and $q_5$ are control qubits
and $q_0$ and $q_1$ are target qubits.
The action of $U_2$ on $I_{q_0}\otimes I_{q_1}\otimes|\xi\ra_2$ gives
\begin{align}\label{eq:app_cx_xi2}
   U_2|\xi\ra_2
  =\dfrac{1}{\sqrt{2}}
  \Big(
    IX \otimes |\om\!\xm\ra
    + XI \otimes |\xm\!\om\ra
  \Big),
\end{align}
where, again, we omit $I_{q_0}\otimes I_{q_1}$ for short.
After the interaction $U_2$,
we apply the Hadamard gates 
${\rm H}_{q_4}{\rm H}_{q_5}$ onto
qubits $q_4$ and $q_5$ of the meter M$_2$
\begin{widetext}
\begin{align}\label{eq:app_HH_cx_xi2}
  \notag{\rm H}_{q_4}{\rm H}_{q_5}U_2|\xi\ra_2
  &=\dfrac{1}{2\sqrt{2}}
  \Big(
    IX \otimes \big[|\om\ra+|\xm\ra\big)
    \big(|\om\ra-|\xm\ra\big)
    + XI \otimes 
    \big(|\om\ra - |\xm\ra\big)
    \big(|\om\ra + |\xm\ra\big)
  \Big]\\
  &=\dfrac{1}{2\sqrt{2}}
  \Big[
  \big(IX + XI\big)\otimes\big(|\om\!\om\ra-|\xm\!\xm\ra\big)
  +\big(IX - XI\big)
  \otimes\big(-|\om\!\xm\ra+|\xm\!\om\ra\big)
  \Big],
\end{align}
\end{widetext}
where we used ${\rm H}|\om\ra = (|\om\ra + |\xm\ra)/\sqrt{2}$
and ${\rm H}|\xm\ra = (|\om\ra - |\xm\ra)/\sqrt{2}$.
We next calculate the Kraus operators
$N_{\nu} = \la \nu|{\rm H}_{q_4}{\rm H}_{q_5} U_2|\xi\ra_2$, 
where $\nu\in\{\om\om, \om\xm, \xm\om, \xm\xm\}$.
We have
\begin{align}
    N_{\om\om} &= -N_{\xm\xm} = \dfrac{1}{2\sqrt{2}}
    \big(IX + XI\big),\\
    -N_{\om\xm} &= N_{\xm\om} = \dfrac{1}{2\sqrt{2}}
    \big(IX - XI\big),
\end{align}
and the corresponding POVM yields
\begin{align}
    E_{\om\om} &= E_{\xm\xm} = \dfrac{1}{4}\big(II+XX\big), \label{eq:appEoo}\\
    E_{\om\xm} &= E_{\xm\om} = \dfrac{1}{4}\big(II-XX\big), \label{eq:appEox}
\end{align}
which satisfies $\sum_\nu E_{\nu} = II$.
The probabilities yield
\begin{align}
    P_{\om\om} &= P_{\xm\!\xm} 
    = \la\psi|E_{\om\om}|\psi\ra
    = \dfrac{1}{4}\big(1+\cos\phi\sin2\theta\big),\\
    P_{\om\xm} &= P_{\xm\om} 
    = \la\psi|E_{\om\xm}|\psi\ra
    = \dfrac{1}{4}\big(1-\cos\phi\sin2\theta\big).
\end{align}
Finally, we get the expectation value $\la XX\ra$
\begin{align}\label{eq:app_XX}
   \la XX\ra =  P_{\om\om} - P_{\om\xm} - P_{\xm\om} + P_{\xm\!\xm} = 
    \cos\phi\sin2\theta,
\end{align}
as shown in Eq.~\eqref{eq:XX} in the main text.

\setcounter{equation}{0}
\renewcommand{\theequation}{E.\arabic{equation}}
\section{Post-measurement state}
\label{appE}
In this section, we derive the final system state after measuring M$_1$ and M$_2$.
The system state after the first measurement gives $\rho_1 = \sum_\mu M_\mu|\psi\ra\la\psi|M_\mu^\dagger$.
We first derive
\begin{align}\label{eq:app_Mupuppsi}
\notag M_{\up\up}|\psi\ra &= 
\dfrac{1}{\sqrt{2}}\big(|00\ra\la00|+|11\ra\la11|\big)
\big(\cos\theta|00\ra + e^{i\phi}\sin\theta|11\ra\big)\\
&= \dfrac{1}{\sqrt{2}}|\psi\ra,
\end{align}
and thus $M_{\up\up}|\psi\ra\la\psi|M^\dagger_{\up\up} = |\psi\ra/2$.
Similarly, we have $M_{\dn\dn}|\psi\ra\la\psi|M^\dagger_{\dn\dn} = |\psi\ra/2$, and
$M_{\up\dn}|\psi\ra\la\psi|M^\dagger_{\up\dn} = M_{\dn\up}|\psi\ra\la\psi|M^\dagger_{\dn\up} = 0.$
Finally, we get    
\begin{align}\label{eq:apppsi1}
\rho_1 = \sum_\mu 
    M_\mu|\psi\ra\la\psi|M_\mu^\dagger
    = |\psi\ra\la\psi|,
\end{align}

Next, we derive the system state 
after the second measurement. 
It gives $\rho_2 = \sum_\nu 
    N_\nu\rho_1N_\nu^\dagger$.
We first derive
\begin{align}\label{eq:app_Noopsi}
\notag N_{\om\om}|\psi\ra &= 
\dfrac{1}{2\sqrt{2}}\big(IX + XI\big)
\big(\cos\theta|00\ra + e^{i\phi}\sin\theta|11\ra\big)\\
&= \dfrac{1}{2\sqrt{2}}
\big(\cos\theta + e^{i\phi}\sin\theta\big)
\big(|01\ra+|10\ra\big),
\end{align}
and thus 
    
\begin{align}\label{eq:appPsi2}
    \rho_2 = \sum_\nu 
    N_\nu\rho_1N_\nu^\dagger
    = \dfrac{1}{2}\begin{pmatrix}
0 & 0 & 0 & 0 \\
0 & 1 & \cos\phi\sin2\theta & 0 \\
0 & \cos\phi\sin2\theta & 1 & 0 \\
0 & 0 & 0 & 0
\end{pmatrix}.
\end{align}

\setcounter{equation}{0}
\renewcommand{\theequation}{F.\arabic{equation}}
\section{Mixed state case}
\label{appF}
In this section, we provide detailed calculations for mixed-state cases. 
We need to derive Eqs.(\ref{eq:mix_Puu}-\ref{eq:mix_Pox}).
We first recast the quantum state in matrix form as
\begin{align}\label{eq:app_rho}
    \rho = 
\begin{pmatrix}
\frac{1 - p}{4} & 0 & 0 & 0 \\
0 & \frac{1 + p}{4} & -\frac{p}{2} & 0 \\
0 & -\frac{p}{2} & \frac{1 + p}{4} & 0 \\
0 & 0 & 0 & \frac{1 - p}{4} 
\end{pmatrix}.
\end{align}
The probabilities give
\begin{align}
    \notag P_{\up\up} &= {\rm Tr}\big[E_{\up\up}\rho\big]\\
    \notag &=\dfrac{1}{2} {\rm Tr}\big[\big(|00\ra\la00|+|11\ra\la11\big)\rho\big]\\
    \notag &=\dfrac{1}{2} \big[\la00|\rho|00\ra + \la11|\rho|11\ra\big]\\
    &= \dfrac{1-p}{4}.
\end{align}
Similarly, we have $P_{\dn\dn} = \dfrac{1-p}{4}$, and 
\begin{align}
\notag P_{\up\dn} = P_{\dn\up} &= \dfrac{1}{2}(\la 01|\rho|01\ra + \la 10|\rho|10\ra)\\
&= \dfrac{1+p}{4}.
\end{align}
Finally,  the expectation value for the measurement of $\la ZZ \ra$ gives

\begin{align}\label{eq:appZZ1}
\notag \la ZZ\ra &= P_{\up\up} - P_{\up\dn} - P_{\dn\up} + P_{\dn\dn} \\
\notag &= \dfrac{1-p}{4} - \dfrac{1+p}{4} - \dfrac{1+p}{4} + \dfrac{1-p}{4} \\
&= -p.
\end{align}

We can also take the expectation value for the measurement of $\la$XX$\ra$. 
First, we calculate the probabilities 
\begin{align}
    \notag P_{\om\om} = P_{\xm\xm} &= {\rm Tr}\big[E_{\om\om}\rho\big]\\
    \notag &=\dfrac{1}{4}{\rm Tr}\big[(II + XX)\rho\big]\\
    &=\dfrac{1-p}{4}, 
\end{align}
\begin{align}
    \notag P_{\om\xm} = P_{\xm\om} &= {\rm Tr}\big[E_{\om\xm}\rho\big]\\
    \notag &=\dfrac{1}{4}{\rm Tr}\big[(II + XX)\rho\big]\\
    &=\dfrac{1+p}{4}
\end{align}
Then, the expectation value
\begin{align}\label{eq:appXX}
\la XX\ra = -p.
\end{align}

The post-measurement states are given through 
$\rho_1 = \sum_\mu 
    M_\mu\rho M_\mu^\dagger$
and 
$\rho_2 = \sum_\nu 
    N_\nu\rho_1N_\nu^\dagger$
after measuring of M$_1$ and M$_2$,
respectively. 
We first derive $M_\mu\rho M_\mu^\dagger$
for $\mu = \{\up\up, \up\dn, \dn\up, \dn\dn\}$, 
where
\begin{align}\label{eq:appupuprho} 
    M_{\up\up}\rho M_{\up\up}^\dagger
    = \dfrac{1}{2}
\begin{pmatrix}
\frac{1 - p}{4} & 0 & 0 & 0 \\
0 & 0 & 0 & 0 \\
0 & 0 & 0 & 0 \\
0 & 0 & 0 & \frac{1 - p}{4} 
\end{pmatrix},
\end{align}
and similar for $M_{\dn\dn}\rho M_{\dn\dn}^\dagger$, and 
\begin{align}\label{eq:appUpDnrho} 
    M_{\up\dn}\rho M_{\up\dn}^\dagger
    = \dfrac{1}{2}
\begin{pmatrix}
0 & 0 & 0 & 0 \\
0 & \frac{1 + p}{4} & -\frac{p}{2} & 0 \\
0 & -\frac{p}{2} & \frac{1 + p}{4} & 0 \\
0 & 0 & 0 & 0
\end{pmatrix},
\end{align}
and similar for $M_{\dn\up}\rho M_{\dn\up}^\dagger$.
Finally, we have
    
\begin{align}\label{eq:appwerner1}
    \rho_1 = \sum_\mu 
    M_\mu\rho M_\mu^\dagger
    = \rho.
\end{align}
Similar for $\rho_2$, we obtain
\begin{align}\label{eq:werner2}
    \rho_2 = \sum_\nu 
    N_\nu\rho_1N_\nu^\dagger
    = \begin{pmatrix}
\frac{1 + p}{4} & 0 & 0 & -\frac{p}{2} \\
0 & \frac{1 - p}{4} & 0 & 0 \\
0 & 0 & \frac{1 - p}{4} & 0 \\
-\frac{p}{2} & 0 & 0 & \frac{1 + p}{4}
\end{pmatrix}.
\end{align}

\bibliography{refs}

\begin{thebibliography}{49}%
\makeatletter
\providecommand \@ifxundefined [1]{%
 \@ifx{#1\undefined}
}%
\providecommand \@ifnum [1]{%
 \ifnum #1\expandafter \@firstoftwo
 \else \expandafter \@secondoftwo
 \fi
}%
\providecommand \@ifx [1]{%
 \ifx #1\expandafter \@firstoftwo
 \else \expandafter \@secondoftwo
 \fi
}%
\providecommand \natexlab [1]{#1}%
\providecommand \enquote  [1]{``#1''}%
\providecommand \bibnamefont  [1]{#1}%
\providecommand \bibfnamefont [1]{#1}%
\providecommand \citenamefont [1]{#1}%
\providecommand \href@noop [0]{\@secondoftwo}%
\providecommand \href [0]{\begingroup \@sanitize@url \@href}%
\providecommand \@href[1]{\@@startlink{#1}\@@href}%
\providecommand \@@href[1]{\endgroup#1\@@endlink}%
\providecommand \@sanitize@url [0]{\catcode `\\12\catcode `\$12\catcode
  `\&12\catcode `\#12\catcode `\^12\catcode `\_12\catcode `\%12\relax}%
\providecommand \@@startlink[1]{}%
\providecommand \@@endlink[0]{}%
\providecommand \url  [0]{\begingroup\@sanitize@url \@url }%
\providecommand \@url [1]{\endgroup\@href {#1}{\urlprefix }}%
\providecommand \urlprefix  [0]{URL }%
\providecommand \Eprint [0]{\href }%
\providecommand \doibase [0]{https://doi.org/}%
\providecommand \selectlanguage [0]{\@gobble}%
\providecommand \bibinfo  [0]{\@secondoftwo}%
\providecommand \bibfield  [0]{\@secondoftwo}%
\providecommand \translation [1]{[#1]}%
\providecommand \BibitemOpen [0]{}%
\providecommand \bibitemStop [0]{}%
\providecommand \bibitemNoStop [0]{.\EOS\space}%
\providecommand \EOS [0]{\spacefactor3000\relax}%
\providecommand \BibitemShut  [1]{\csname bibitem#1\endcsname}%
\let\auto@bib@innerbib\@empty
%</preamble>
\bibitem [{\citenamefont {Horodecki}\ \emph {et~al.}(2009)\citenamefont
  {Horodecki}, \citenamefont {Horodecki}, \citenamefont {Horodecki},\ and\
  \citenamefont {Horodecki}}]{RevModPhys.81.865}%
  \BibitemOpen
  \bibfield  {author} {\bibinfo {author} {\bibfnamefont {R.}~\bibnamefont
  {Horodecki}}, \bibinfo {author} {\bibfnamefont {P.}~\bibnamefont
  {Horodecki}}, \bibinfo {author} {\bibfnamefont {M.}~\bibnamefont
  {Horodecki}},\ and\ \bibinfo {author} {\bibfnamefont {K.}~\bibnamefont
  {Horodecki}},\ }\bibfield  {title} {\bibinfo {title} {Quantum entanglement},\
  }\href {https://doi.org/10.1103/RevModPhys.81.865} {\bibfield  {journal}
  {\bibinfo  {journal} {Rev. Mod. Phys.}\ }\textbf {\bibinfo {volume} {81}},\
  \bibinfo {pages} {865} (\bibinfo {year} {2009})}\BibitemShut {NoStop}%
\bibitem [{\citenamefont {Duarte}(2022)}]{10.1088/978-0-7503-5269-7}%
  \BibitemOpen
  \bibfield  {author} {\bibinfo {author} {\bibfnamefont {F.~J.}\ \bibnamefont
  {Duarte}},\ }\href {https://doi.org/10.1088/978-0-7503-5269-7} {\emph
  {\bibinfo {title} {Fundamentals of Quantum Entanglement (Second Edition)}}},\
  2053-2563\ (\bibinfo  {publisher} {IOP Publishing},\ \bibinfo {year}
  {2022})\BibitemShut {NoStop}%
\bibitem [{\citenamefont {Yu}(2021)}]{Yu_2021}%
  \BibitemOpen
  \bibfield  {author} {\bibinfo {author} {\bibfnamefont {Y.}~\bibnamefont
  {Yu}},\ }\bibfield  {title} {\bibinfo {title} {Advancements in applications
  of quantum entanglement},\ }\href
  {https://doi.org/10.1088/1742-6596/2012/1/012113} {\bibfield  {journal}
  {\bibinfo  {journal} {Journal of Physics: Conference Series}\ }\textbf
  {\bibinfo {volume} {2012}},\ \bibinfo {pages} {012113} (\bibinfo {year}
  {2021})}\BibitemShut {NoStop}%
\bibitem [{\citenamefont {Yin}\ \emph {et~al.}(2020)\citenamefont {Yin},
  \citenamefont {Li}, \citenamefont {Liao}, \citenamefont {Yang}, \citenamefont
  {Cao}, \citenamefont {Zhang}, \citenamefont {Ren}, \citenamefont {Cai},
  \citenamefont {Liu}, \citenamefont {Li}, \citenamefont {Shu}, \citenamefont
  {Huang}, \citenamefont {Deng}, \citenamefont {Li}, \citenamefont {Zhang},
  \citenamefont {Liu}, \citenamefont {Chen}, \citenamefont {Lu}, \citenamefont
  {Wang}, \citenamefont {Xu}, \citenamefont {Wang}, \citenamefont {Peng},
  \citenamefont {Ekert},\ and\ \citenamefont {Pan}}]{Yin2020}%
  \BibitemOpen
  \bibfield  {author} {\bibinfo {author} {\bibfnamefont {J.}~\bibnamefont
  {Yin}}, \bibinfo {author} {\bibfnamefont {Y.-H.}\ \bibnamefont {Li}},
  \bibinfo {author} {\bibfnamefont {S.-K.}\ \bibnamefont {Liao}}, \bibinfo
  {author} {\bibfnamefont {M.}~\bibnamefont {Yang}}, \bibinfo {author}
  {\bibfnamefont {Y.}~\bibnamefont {Cao}}, \bibinfo {author} {\bibfnamefont
  {L.}~\bibnamefont {Zhang}}, \bibinfo {author} {\bibfnamefont {J.-G.}\
  \bibnamefont {Ren}}, \bibinfo {author} {\bibfnamefont {W.-Q.}\ \bibnamefont
  {Cai}}, \bibinfo {author} {\bibfnamefont {W.-Y.}\ \bibnamefont {Liu}},
  \bibinfo {author} {\bibfnamefont {S.-L.}\ \bibnamefont {Li}}, \bibinfo
  {author} {\bibfnamefont {R.}~\bibnamefont {Shu}}, \bibinfo {author}
  {\bibfnamefont {Y.-M.}\ \bibnamefont {Huang}}, \bibinfo {author}
  {\bibfnamefont {L.}~\bibnamefont {Deng}}, \bibinfo {author} {\bibfnamefont
  {L.}~\bibnamefont {Li}}, \bibinfo {author} {\bibfnamefont {Q.}~\bibnamefont
  {Zhang}}, \bibinfo {author} {\bibfnamefont {N.-L.}\ \bibnamefont {Liu}},
  \bibinfo {author} {\bibfnamefont {Y.-A.}\ \bibnamefont {Chen}}, \bibinfo
  {author} {\bibfnamefont {C.-Y.}\ \bibnamefont {Lu}}, \bibinfo {author}
  {\bibfnamefont {X.-B.}\ \bibnamefont {Wang}}, \bibinfo {author}
  {\bibfnamefont {F.}~\bibnamefont {Xu}}, \bibinfo {author} {\bibfnamefont
  {J.-Y.}\ \bibnamefont {Wang}}, \bibinfo {author} {\bibfnamefont {C.-Z.}\
  \bibnamefont {Peng}}, \bibinfo {author} {\bibfnamefont {A.~K.}\ \bibnamefont
  {Ekert}},\ and\ \bibinfo {author} {\bibfnamefont {J.-W.}\ \bibnamefont
  {Pan}},\ }\bibfield  {title} {\bibinfo {title} {Entanglement-based secure
  quantum cryptography over 1,120 kilometres},\ }\href
  {https://doi.org/10.1038/s41586-020-2401-y} {\bibfield  {journal} {\bibinfo
  {journal} {Nature}\ }\textbf {\bibinfo {volume} {582}},\ \bibinfo {pages}
  {501} (\bibinfo {year} {2020})}\BibitemShut {NoStop}%
\bibitem [{\citenamefont {Basset}\ \emph {et~al.}(2021)\citenamefont {Basset},
  \citenamefont {Valeri}, \citenamefont {Roccia}, \citenamefont {Muredda},
  \citenamefont {Poderini}, \citenamefont {Neuwirth}, \citenamefont {Spagnolo},
  \citenamefont {Rota}, \citenamefont {Carvacho}, \citenamefont {Sciarrino},\
  and\ \citenamefont {Trotta}}]{doi:10.1126/sciadv.abe6379}%
  \BibitemOpen
  \bibfield  {author} {\bibinfo {author} {\bibfnamefont {F.~B.}\ \bibnamefont
  {Basset}}, \bibinfo {author} {\bibfnamefont {M.}~\bibnamefont {Valeri}},
  \bibinfo {author} {\bibfnamefont {E.}~\bibnamefont {Roccia}}, \bibinfo
  {author} {\bibfnamefont {V.}~\bibnamefont {Muredda}}, \bibinfo {author}
  {\bibfnamefont {D.}~\bibnamefont {Poderini}}, \bibinfo {author}
  {\bibfnamefont {J.}~\bibnamefont {Neuwirth}}, \bibinfo {author}
  {\bibfnamefont {N.}~\bibnamefont {Spagnolo}}, \bibinfo {author}
  {\bibfnamefont {M.~B.}\ \bibnamefont {Rota}}, \bibinfo {author}
  {\bibfnamefont {G.}~\bibnamefont {Carvacho}}, \bibinfo {author}
  {\bibfnamefont {F.}~\bibnamefont {Sciarrino}},\ and\ \bibinfo {author}
  {\bibfnamefont {R.}~\bibnamefont {Trotta}},\ }\bibfield  {title} {\bibinfo
  {title} {Quantum key distribution with entangled photons generated on demand
  by a quantum dot},\ }\href {https://doi.org/10.1126/sciadv.abe6379}
  {\bibfield  {journal} {\bibinfo  {journal} {Science Advances}\ }\textbf
  {\bibinfo {volume} {7}},\ \bibinfo {pages} {eabe6379} (\bibinfo {year}
  {2021})},\ \Eprint
  {https://arxiv.org/abs/https://www.science.org/doi/pdf/10.1126/sciadv.abe6379}
  {https://www.science.org/doi/pdf/10.1126/sciadv.abe6379} \BibitemShut
  {NoStop}%
\bibitem [{\citenamefont {Zou}(2021)}]{Zou_2021}%
  \BibitemOpen
  \bibfield  {author} {\bibinfo {author} {\bibfnamefont {N.}~\bibnamefont
  {Zou}},\ }\bibfield  {title} {\bibinfo {title} {Quantum entanglement and its
  application in quantum communication},\ }\href
  {https://doi.org/10.1088/1742-6596/1827/1/012120} {\bibfield  {journal}
  {\bibinfo  {journal} {Journal of Physics: Conference Series}\ }\textbf
  {\bibinfo {volume} {1827}},\ \bibinfo {pages} {012120} (\bibinfo {year}
  {2021})}\BibitemShut {NoStop}%
\bibitem [{\citenamefont {Wengerowsky}\ \emph {et~al.}(2018)\citenamefont
  {Wengerowsky}, \citenamefont {Joshi}, \citenamefont {Steinlechner},
  \citenamefont {H{\"u}bel},\ and\ \citenamefont {Ursin}}]{Wengerowsky2018}%
  \BibitemOpen
  \bibfield  {author} {\bibinfo {author} {\bibfnamefont {S.}~\bibnamefont
  {Wengerowsky}}, \bibinfo {author} {\bibfnamefont {S.~K.}\ \bibnamefont
  {Joshi}}, \bibinfo {author} {\bibfnamefont {F.}~\bibnamefont {Steinlechner}},
  \bibinfo {author} {\bibfnamefont {H.}~\bibnamefont {H{\"u}bel}},\ and\
  \bibinfo {author} {\bibfnamefont {R.}~\bibnamefont {Ursin}},\ }\bibfield
  {title} {\bibinfo {title} {An entanglement-based wavelength-multiplexed
  quantum communication network},\ }\href
  {https://doi.org/10.1038/s41586-018-0766-y} {\bibfield  {journal} {\bibinfo
  {journal} {Nature}\ }\textbf {\bibinfo {volume} {564}},\ \bibinfo {pages}
  {225} (\bibinfo {year} {2018})}\BibitemShut {NoStop}%
\bibitem [{\citenamefont {Guccione}\ \emph {et~al.}(2020)\citenamefont
  {Guccione}, \citenamefont {Darras}, \citenamefont {Jeannic}, \citenamefont
  {Verma}, \citenamefont {Nam}, \citenamefont {Cavaillès},\ and\ \citenamefont
  {Laurat}}]{doi:10.1126/sciadv.aba4508}%
  \BibitemOpen
  \bibfield  {author} {\bibinfo {author} {\bibfnamefont {G.}~\bibnamefont
  {Guccione}}, \bibinfo {author} {\bibfnamefont {T.}~\bibnamefont {Darras}},
  \bibinfo {author} {\bibfnamefont {H.~L.}\ \bibnamefont {Jeannic}}, \bibinfo
  {author} {\bibfnamefont {V.~B.}\ \bibnamefont {Verma}}, \bibinfo {author}
  {\bibfnamefont {S.~W.}\ \bibnamefont {Nam}}, \bibinfo {author} {\bibfnamefont
  {A.}~\bibnamefont {Cavaillès}},\ and\ \bibinfo {author} {\bibfnamefont
  {J.}~\bibnamefont {Laurat}},\ }\bibfield  {title} {\bibinfo {title}
  {Connecting heterogeneous quantum networks by hybrid entanglement swapping},\
  }\href {https://doi.org/10.1126/sciadv.aba4508} {\bibfield  {journal}
  {\bibinfo  {journal} {Science Advances}\ }\textbf {\bibinfo {volume} {6}},\
  \bibinfo {pages} {eaba4508} (\bibinfo {year} {2020})},\ \Eprint
  {https://arxiv.org/abs/https://www.science.org/doi/pdf/10.1126/sciadv.aba4508}
  {https://www.science.org/doi/pdf/10.1126/sciadv.aba4508} \BibitemShut
  {NoStop}%
\bibitem [{\citenamefont {Huang}\ \emph {et~al.}(2024)\citenamefont {Huang},
  \citenamefont {Zhuang},\ and\ \citenamefont {Lee}}]{10.1063/5.0204102}%
  \BibitemOpen
  \bibfield  {author} {\bibinfo {author} {\bibfnamefont {J.}~\bibnamefont
  {Huang}}, \bibinfo {author} {\bibfnamefont {M.}~\bibnamefont {Zhuang}},\ and\
  \bibinfo {author} {\bibfnamefont {C.}~\bibnamefont {Lee}},\ }\bibfield
  {title} {\bibinfo {title} {{Entanglement-enhanced quantum metrology: From
  standard quantum limit to Heisenberg limit}},\ }\href
  {https://doi.org/10.1063/5.0204102} {\bibfield  {journal} {\bibinfo
  {journal} {Applied Physics Reviews}\ }\textbf {\bibinfo {volume} {11}},\
  \bibinfo {pages} {031302} (\bibinfo {year} {2024})},\ \Eprint
  {https://arxiv.org/abs/https://pubs.aip.org/aip/apr/article-pdf/doi/10.1063/5.0204102/20027645/031302\_1\_5.0204102.pdf}
  {https://pubs.aip.org/aip/apr/article-pdf/doi/10.1063/5.0204102/20027645/031302\_1\_5.0204102.pdf}
  \BibitemShut {NoStop}%
\bibitem [{\citenamefont {Augusiak}\ \emph {et~al.}(2016)\citenamefont
  {Augusiak}, \citenamefont {Ko\l{}ody\ifmmode~\acute{n}\else \'{n}\fi{}ski},
  \citenamefont {Streltsov}, \citenamefont {Bera}, \citenamefont {Ac\'{\i}n},\
  and\ \citenamefont {Lewenstein}}]{PhysRevA.94.012339}%
  \BibitemOpen
  \bibfield  {author} {\bibinfo {author} {\bibfnamefont {R.}~\bibnamefont
  {Augusiak}}, \bibinfo {author} {\bibfnamefont {J.}~\bibnamefont
  {Ko\l{}ody\ifmmode~\acute{n}\else \'{n}\fi{}ski}}, \bibinfo {author}
  {\bibfnamefont {A.}~\bibnamefont {Streltsov}}, \bibinfo {author}
  {\bibfnamefont {M.~N.}\ \bibnamefont {Bera}}, \bibinfo {author}
  {\bibfnamefont {A.}~\bibnamefont {Ac\'{\i}n}},\ and\ \bibinfo {author}
  {\bibfnamefont {M.}~\bibnamefont {Lewenstein}},\ }\bibfield  {title}
  {\bibinfo {title} {Asymptotic role of entanglement in quantum metrology},\
  }\href {https://doi.org/10.1103/PhysRevA.94.012339} {\bibfield  {journal}
  {\bibinfo  {journal} {Phys. Rev. A}\ }\textbf {\bibinfo {volume} {94}},\
  \bibinfo {pages} {012339} (\bibinfo {year} {2016})}\BibitemShut {NoStop}%
\bibitem [{\citenamefont {Gühne}\ and\ \citenamefont
  {Tóth}(2009)}]{GUHNE20091}%
  \BibitemOpen
  \bibfield  {author} {\bibinfo {author} {\bibfnamefont {O.}~\bibnamefont
  {Gühne}}\ and\ \bibinfo {author} {\bibfnamefont {G.}~\bibnamefont {Tóth}},\
  }\bibfield  {title} {\bibinfo {title} {Entanglement detection},\ }\href
  {https://doi.org/https://doi.org/10.1016/j.physrep.2009.02.004} {\bibfield
  {journal} {\bibinfo  {journal} {Physics Reports}\ }\textbf {\bibinfo {volume}
  {474}},\ \bibinfo {pages} {1} (\bibinfo {year} {2009})}\BibitemShut {NoStop}%
\bibitem [{\citenamefont {Gurvits}(2004)}]{GURVITS2004448}%
  \BibitemOpen
  \bibfield  {author} {\bibinfo {author} {\bibfnamefont {L.}~\bibnamefont
  {Gurvits}},\ }\bibfield  {title} {\bibinfo {title} {Classical complexity and
  quantum entanglement},\ }\href
  {https://doi.org/https://doi.org/10.1016/j.jcss.2004.06.003} {\bibfield
  {journal} {\bibinfo  {journal} {Journal of Computer and System Sciences}\
  }\textbf {\bibinfo {volume} {69}},\ \bibinfo {pages} {448} (\bibinfo {year}
  {2004})},\ \bibinfo {note} {special Issue on STOC 2003}\BibitemShut {NoStop}%
\bibitem [{\citenamefont {James}\ \emph {et~al.}(2001)\citenamefont {James},
  \citenamefont {Kwiat}, \citenamefont {Munro},\ and\ \citenamefont
  {White}}]{PhysRevA.64.052312}%
  \BibitemOpen
  \bibfield  {author} {\bibinfo {author} {\bibfnamefont {D.~F.~V.}\
  \bibnamefont {James}}, \bibinfo {author} {\bibfnamefont {P.~G.}\ \bibnamefont
  {Kwiat}}, \bibinfo {author} {\bibfnamefont {W.~J.}\ \bibnamefont {Munro}},\
  and\ \bibinfo {author} {\bibfnamefont {A.~G.}\ \bibnamefont {White}},\
  }\bibfield  {title} {\bibinfo {title} {Measurement of qubits},\ }\href
  {https://doi.org/10.1103/PhysRevA.64.052312} {\bibfield  {journal} {\bibinfo
  {journal} {Phys. Rev. A}\ }\textbf {\bibinfo {volume} {64}},\ \bibinfo
  {pages} {052312} (\bibinfo {year} {2001})}\BibitemShut {NoStop}%
\bibitem [{\citenamefont {Thew}\ \emph {et~al.}(2002)\citenamefont {Thew},
  \citenamefont {Nemoto}, \citenamefont {White},\ and\ \citenamefont
  {Munro}}]{PhysRevA.66.012303}%
  \BibitemOpen
  \bibfield  {author} {\bibinfo {author} {\bibfnamefont {R.~T.}\ \bibnamefont
  {Thew}}, \bibinfo {author} {\bibfnamefont {K.}~\bibnamefont {Nemoto}},
  \bibinfo {author} {\bibfnamefont {A.~G.}\ \bibnamefont {White}},\ and\
  \bibinfo {author} {\bibfnamefont {W.~J.}\ \bibnamefont {Munro}},\ }\bibfield
  {title} {\bibinfo {title} {Qudit quantum-state tomography},\ }\href
  {https://doi.org/10.1103/PhysRevA.66.012303} {\bibfield  {journal} {\bibinfo
  {journal} {Phys. Rev. A}\ }\textbf {\bibinfo {volume} {66}},\ \bibinfo
  {pages} {012303} (\bibinfo {year} {2002})}\BibitemShut {NoStop}%
\bibitem [{\citenamefont {Peres}(1996)}]{PhysRevLett.77.1413}%
  \BibitemOpen
  \bibfield  {author} {\bibinfo {author} {\bibfnamefont {A.}~\bibnamefont
  {Peres}},\ }\bibfield  {title} {\bibinfo {title} {Separability criterion for
  density matrices},\ }\href {https://doi.org/10.1103/PhysRevLett.77.1413}
  {\bibfield  {journal} {\bibinfo  {journal} {Phys. Rev. Lett.}\ }\textbf
  {\bibinfo {volume} {77}},\ \bibinfo {pages} {1413} (\bibinfo {year}
  {1996})}\BibitemShut {NoStop}%
\bibitem [{\citenamefont {Horodecki}\ \emph {et~al.}(1996)\citenamefont
  {Horodecki}, \citenamefont {Horodecki},\ and\ \citenamefont
  {Horodecki}}]{HORODECKI19961}%
  \BibitemOpen
  \bibfield  {author} {\bibinfo {author} {\bibfnamefont {M.}~\bibnamefont
  {Horodecki}}, \bibinfo {author} {\bibfnamefont {P.}~\bibnamefont
  {Horodecki}},\ and\ \bibinfo {author} {\bibfnamefont {R.}~\bibnamefont
  {Horodecki}},\ }\bibfield  {title} {\bibinfo {title} {Separability of mixed
  states: necessary and sufficient conditions},\ }\href
  {https://doi.org/https://doi.org/10.1016/S0375-9601(96)00706-2} {\bibfield
  {journal} {\bibinfo  {journal} {Physics Letters A}\ }\textbf {\bibinfo
  {volume} {223}},\ \bibinfo {pages} {1} (\bibinfo {year} {1996})}\BibitemShut
  {NoStop}%
\bibitem [{\citenamefont {Hill}\ and\ \citenamefont
  {Wootters}(1997)}]{PhysRevLett.78.5022}%
  \BibitemOpen
  \bibfield  {author} {\bibinfo {author} {\bibfnamefont {S.~A.}\ \bibnamefont
  {Hill}}\ and\ \bibinfo {author} {\bibfnamefont {W.~K.}\ \bibnamefont
  {Wootters}},\ }\bibfield  {title} {\bibinfo {title} {Entanglement of a pair
  of quantum bits},\ }\href {https://doi.org/10.1103/PhysRevLett.78.5022}
  {\bibfield  {journal} {\bibinfo  {journal} {Phys. Rev. Lett.}\ }\textbf
  {\bibinfo {volume} {78}},\ \bibinfo {pages} {5022} (\bibinfo {year}
  {1997})}\BibitemShut {NoStop}%
\bibitem [{\citenamefont {Elben}\ \emph {et~al.}(2020)\citenamefont {Elben},
  \citenamefont {Kueng}, \citenamefont {Huang}, \citenamefont {van Bijnen},
  \citenamefont {Kokail}, \citenamefont {Dalmonte}, \citenamefont {Calabrese},
  \citenamefont {Kraus}, \citenamefont {Preskill}, \citenamefont {Zoller},\
  and\ \citenamefont {Vermersch}}]{PhysRevLett.125.200501}%
  \BibitemOpen
  \bibfield  {author} {\bibinfo {author} {\bibfnamefont {A.}~\bibnamefont
  {Elben}}, \bibinfo {author} {\bibfnamefont {R.}~\bibnamefont {Kueng}},
  \bibinfo {author} {\bibfnamefont {H.-Y.~R.}\ \bibnamefont {Huang}}, \bibinfo
  {author} {\bibfnamefont {R.}~\bibnamefont {van Bijnen}}, \bibinfo {author}
  {\bibfnamefont {C.}~\bibnamefont {Kokail}}, \bibinfo {author} {\bibfnamefont
  {M.}~\bibnamefont {Dalmonte}}, \bibinfo {author} {\bibfnamefont
  {P.}~\bibnamefont {Calabrese}}, \bibinfo {author} {\bibfnamefont
  {B.}~\bibnamefont {Kraus}}, \bibinfo {author} {\bibfnamefont
  {J.}~\bibnamefont {Preskill}}, \bibinfo {author} {\bibfnamefont
  {P.}~\bibnamefont {Zoller}},\ and\ \bibinfo {author} {\bibfnamefont
  {B.}~\bibnamefont {Vermersch}},\ }\bibfield  {title} {\bibinfo {title}
  {Mixed-state entanglement from local randomized measurements},\ }\href
  {https://doi.org/10.1103/PhysRevLett.125.200501} {\bibfield  {journal}
  {\bibinfo  {journal} {Phys. Rev. Lett.}\ }\textbf {\bibinfo {volume} {125}},\
  \bibinfo {pages} {200501} (\bibinfo {year} {2020})}\BibitemShut {NoStop}%
\bibitem [{\citenamefont {Bourennane}\ \emph {et~al.}(2004)\citenamefont
  {Bourennane}, \citenamefont {Eibl}, \citenamefont {Kurtsiefer}, \citenamefont
  {Gaertner}, \citenamefont {Weinfurter}, \citenamefont {G\"uhne},
  \citenamefont {Hyllus}, \citenamefont {Bru\ss{}}, \citenamefont
  {Lewenstein},\ and\ \citenamefont {Sanpera}}]{PhysRevLett.92.087902}%
  \BibitemOpen
  \bibfield  {author} {\bibinfo {author} {\bibfnamefont {M.}~\bibnamefont
  {Bourennane}}, \bibinfo {author} {\bibfnamefont {M.}~\bibnamefont {Eibl}},
  \bibinfo {author} {\bibfnamefont {C.}~\bibnamefont {Kurtsiefer}}, \bibinfo
  {author} {\bibfnamefont {S.}~\bibnamefont {Gaertner}}, \bibinfo {author}
  {\bibfnamefont {H.}~\bibnamefont {Weinfurter}}, \bibinfo {author}
  {\bibfnamefont {O.}~\bibnamefont {G\"uhne}}, \bibinfo {author} {\bibfnamefont
  {P.}~\bibnamefont {Hyllus}}, \bibinfo {author} {\bibfnamefont
  {D.}~\bibnamefont {Bru\ss{}}}, \bibinfo {author} {\bibfnamefont
  {M.}~\bibnamefont {Lewenstein}},\ and\ \bibinfo {author} {\bibfnamefont
  {A.}~\bibnamefont {Sanpera}},\ }\bibfield  {title} {\bibinfo {title}
  {Experimental detection of multipartite entanglement using witness
  operators},\ }\href {https://doi.org/10.1103/PhysRevLett.92.087902}
  {\bibfield  {journal} {\bibinfo  {journal} {Phys. Rev. Lett.}\ }\textbf
  {\bibinfo {volume} {92}},\ \bibinfo {pages} {087902} (\bibinfo {year}
  {2004})}\BibitemShut {NoStop}%
\bibitem [{\citenamefont {Tr\'avn\'{\i}\ifmmode~\check{c}\else \v{c}\fi{}ek}\
  \emph {et~al.}(2024)\citenamefont {Tr\'avn\'{\i}\ifmmode~\check{c}\else
  \v{c}\fi{}ek}, \citenamefont {Roik}, \citenamefont {Bartkiewicz},
  \citenamefont {\ifmmode~\check{C}\else \v{C}\fi{}ernoch}, \citenamefont
  {Horodecki},\ and\ \citenamefont {Lemr}}]{PhysRevResearch.6.033056}%
  \BibitemOpen
  \bibfield  {author} {\bibinfo {author} {\bibfnamefont {V.~c.~v.}\
  \bibnamefont {Tr\'avn\'{\i}\ifmmode~\check{c}\else \v{c}\fi{}ek}}, \bibinfo
  {author} {\bibfnamefont {J.}~\bibnamefont {Roik}}, \bibinfo {author}
  {\bibfnamefont {K.}~\bibnamefont {Bartkiewicz}}, \bibinfo {author}
  {\bibfnamefont {A.}~\bibnamefont {\ifmmode~\check{C}\else \v{C}\fi{}ernoch}},
  \bibinfo {author} {\bibfnamefont {P.}~\bibnamefont {Horodecki}},\ and\
  \bibinfo {author} {\bibfnamefont {K.}~\bibnamefont {Lemr}},\ }\bibfield
  {title} {\bibinfo {title} {Sensitivity versus selectivity in entanglement
  detection via collective witnesses},\ }\href
  {https://doi.org/10.1103/PhysRevResearch.6.033056} {\bibfield  {journal}
  {\bibinfo  {journal} {Phys. Rev. Res.}\ }\textbf {\bibinfo {volume} {6}},\
  \bibinfo {pages} {033056} (\bibinfo {year} {2024})}\BibitemShut {NoStop}%
\bibitem [{\citenamefont {Bae}\ \emph {et~al.}(2020)\citenamefont {Bae},
  \citenamefont {Chru{\'{s}}ci{\'{n}}ski},\ and\ \citenamefont
  {Hiesmayr}}]{Bae2020}%
  \BibitemOpen
  \bibfield  {author} {\bibinfo {author} {\bibfnamefont {J.}~\bibnamefont
  {Bae}}, \bibinfo {author} {\bibfnamefont {D.}~\bibnamefont
  {Chru{\'{s}}ci{\'{n}}ski}},\ and\ \bibinfo {author} {\bibfnamefont {B.~C.}\
  \bibnamefont {Hiesmayr}},\ }\bibfield  {title} {\bibinfo {title} {Mirrored
  entanglement witnesses},\ }\href {https://doi.org/10.1038/s41534-020-0242-z}
  {\bibfield  {journal} {\bibinfo  {journal} {npj Quantum Information}\
  }\textbf {\bibinfo {volume} {6}},\ \bibinfo {pages} {15} (\bibinfo {year}
  {2020})}\BibitemShut {NoStop}%
\bibitem [{\citenamefont {Siudzi{\'{n}}ska}\ and\ \citenamefont
  {Chru{\'{s}}ci{\'{n}}ski}(2021)}]{Siudzinska2021}%
  \BibitemOpen
  \bibfield  {author} {\bibinfo {author} {\bibfnamefont {K.}~\bibnamefont
  {Siudzi{\'{n}}ska}}\ and\ \bibinfo {author} {\bibfnamefont {D.}~\bibnamefont
  {Chru{\'{s}}ci{\'{n}}ski}},\ }\bibfield  {title} {\bibinfo {title}
  {Entanglement witnesses from mutually unbiased measurements},\ }\href
  {https://doi.org/10.1038/s41598-021-02356-2} {\bibfield  {journal} {\bibinfo
  {journal} {Scientific Reports}\ }\textbf {\bibinfo {volume} {11}},\ \bibinfo
  {pages} {22988} (\bibinfo {year} {2021})}\BibitemShut {NoStop}%
\bibitem [{\citenamefont {Roik}\ \emph {et~al.}(2021)\citenamefont {Roik},
  \citenamefont {Bartkiewicz}, \citenamefont {\ifmmode~\check{C}\else
  \v{C}\fi{}ernoch},\ and\ \citenamefont {Lemr}}]{PhysRevApplied.15.054006}%
  \BibitemOpen
  \bibfield  {author} {\bibinfo {author} {\bibfnamefont {J.}~\bibnamefont
  {Roik}}, \bibinfo {author} {\bibfnamefont {K.}~\bibnamefont {Bartkiewicz}},
  \bibinfo {author} {\bibfnamefont {A.}~\bibnamefont {\ifmmode~\check{C}\else
  \v{C}\fi{}ernoch}},\ and\ \bibinfo {author} {\bibfnamefont {K.}~\bibnamefont
  {Lemr}},\ }\bibfield  {title} {\bibinfo {title} {Accuracy of entanglement
  detection via artificial neural networks and human-designed entanglement
  witnesses},\ }\href {https://doi.org/10.1103/PhysRevApplied.15.054006}
  {\bibfield  {journal} {\bibinfo  {journal} {Phys. Rev. Appl.}\ }\textbf
  {\bibinfo {volume} {15}},\ \bibinfo {pages} {054006} (\bibinfo {year}
  {2021})}\BibitemShut {NoStop}%
\bibitem [{\citenamefont {Jir{\'a}kov{\'a}}\ \emph {et~al.}(2024)\citenamefont
  {Jir{\'a}kov{\'a}}, \citenamefont {{\v{C}}ernoch}, \citenamefont
  {Barasi{\'{n}}ski},\ and\ \citenamefont {Lemr}}]{Jirakova2024}%
  \BibitemOpen
  \bibfield  {author} {\bibinfo {author} {\bibfnamefont {K.}~\bibnamefont
  {Jir{\'a}kov{\'a}}}, \bibinfo {author} {\bibfnamefont {A.}~\bibnamefont
  {{\v{C}}ernoch}}, \bibinfo {author} {\bibfnamefont {A.}~\bibnamefont
  {Barasi{\'{n}}ski}},\ and\ \bibinfo {author} {\bibfnamefont {K.}~\bibnamefont
  {Lemr}},\ }\bibfield  {title} {\bibinfo {title} {Enhancing collective
  entanglement witnesses through correlation with state purity},\ }\href
  {https://doi.org/10.1038/s41598-024-65385-7} {\bibfield  {journal} {\bibinfo
  {journal} {Scientific Reports}\ }\textbf {\bibinfo {volume} {14}},\ \bibinfo
  {pages} {16374} (\bibinfo {year} {2024})}\BibitemShut {NoStop}%
\bibitem [{\citenamefont {Bell}(1964)}]{PhysicsPhysiqueFizika.1.195}%
  \BibitemOpen
  \bibfield  {author} {\bibinfo {author} {\bibfnamefont {J.~S.}\ \bibnamefont
  {Bell}},\ }\bibfield  {title} {\bibinfo {title} {On the einstein podolsky
  rosen paradox},\ }\href {https://doi.org/10.1103/PhysicsPhysiqueFizika.1.195}
  {\bibfield  {journal} {\bibinfo  {journal} {Physics Physique Fizika}\
  }\textbf {\bibinfo {volume} {1}},\ \bibinfo {pages} {195} (\bibinfo {year}
  {1964})}\BibitemShut {NoStop}%
\bibitem [{\citenamefont {Clauser}\ and\ \citenamefont
  {Shimony}(1978)}]{Clauser_1978}%
  \BibitemOpen
  \bibfield  {author} {\bibinfo {author} {\bibfnamefont {J.~F.}\ \bibnamefont
  {Clauser}}\ and\ \bibinfo {author} {\bibfnamefont {A.}~\bibnamefont
  {Shimony}},\ }\bibfield  {title} {\bibinfo {title} {Bell's theorem.
  experimental tests and implications},\ }\href
  {https://doi.org/10.1088/0034-4885/41/12/002} {\bibfield  {journal} {\bibinfo
   {journal} {Reports on Progress in Physics}\ }\textbf {\bibinfo {volume}
  {41}},\ \bibinfo {pages} {1881} (\bibinfo {year} {1978})}\BibitemShut
  {NoStop}%
\bibitem [{\citenamefont {Clauser}\ \emph {et~al.}(1969)\citenamefont
  {Clauser}, \citenamefont {Horne}, \citenamefont {Shimony},\ and\
  \citenamefont {Holt}}]{PhysRevLett.23.880}%
  \BibitemOpen
  \bibfield  {author} {\bibinfo {author} {\bibfnamefont {J.~F.}\ \bibnamefont
  {Clauser}}, \bibinfo {author} {\bibfnamefont {M.~A.}\ \bibnamefont {Horne}},
  \bibinfo {author} {\bibfnamefont {A.}~\bibnamefont {Shimony}},\ and\ \bibinfo
  {author} {\bibfnamefont {R.~A.}\ \bibnamefont {Holt}},\ }\bibfield  {title}
  {\bibinfo {title} {Proposed experiment to test local hidden-variable
  theories},\ }\href {https://doi.org/10.1103/PhysRevLett.23.880} {\bibfield
  {journal} {\bibinfo  {journal} {Phys. Rev. Lett.}\ }\textbf {\bibinfo
  {volume} {23}},\ \bibinfo {pages} {880} (\bibinfo {year} {1969})}\BibitemShut
  {NoStop}%
\bibitem [{\citenamefont {Bartkiewicz}\ \emph {et~al.}(2013)\citenamefont
  {Bartkiewicz}, \citenamefont {Horst}, \citenamefont {Lemr},\ and\
  \citenamefont {Miranowicz}}]{PhysRevA.88.052105}%
  \BibitemOpen
  \bibfield  {author} {\bibinfo {author} {\bibfnamefont {K.}~\bibnamefont
  {Bartkiewicz}}, \bibinfo {author} {\bibfnamefont {B.}~\bibnamefont {Horst}},
  \bibinfo {author} {\bibfnamefont {K.}~\bibnamefont {Lemr}},\ and\ \bibinfo
  {author} {\bibfnamefont {A.}~\bibnamefont {Miranowicz}},\ }\bibfield  {title}
  {\bibinfo {title} {Entanglement estimation from bell inequality violation},\
  }\href {https://doi.org/10.1103/PhysRevA.88.052105} {\bibfield  {journal}
  {\bibinfo  {journal} {Phys. Rev. A}\ }\textbf {\bibinfo {volume} {88}},\
  \bibinfo {pages} {052105} (\bibinfo {year} {2013})}\BibitemShut {NoStop}%
\bibitem [{\citenamefont {Dong}\ \emph {et~al.}(2023)\citenamefont {Dong},
  \citenamefont {Song}, \citenamefont {Fan}, \citenamefont {Ye},\ and\
  \citenamefont {Wang}}]{PhysRevA.107.052403}%
  \BibitemOpen
  \bibfield  {author} {\bibinfo {author} {\bibfnamefont {D.-D.}\ \bibnamefont
  {Dong}}, \bibinfo {author} {\bibfnamefont {X.-K.}\ \bibnamefont {Song}},
  \bibinfo {author} {\bibfnamefont {X.-G.}\ \bibnamefont {Fan}}, \bibinfo
  {author} {\bibfnamefont {L.}~\bibnamefont {Ye}},\ and\ \bibinfo {author}
  {\bibfnamefont {D.}~\bibnamefont {Wang}},\ }\bibfield  {title} {\bibinfo
  {title} {Complementary relations of entanglement, coherence, steering, and
  bell nonlocality inequality violation in three-qubit states},\ }\href
  {https://doi.org/10.1103/PhysRevA.107.052403} {\bibfield  {journal} {\bibinfo
   {journal} {Phys. Rev. A}\ }\textbf {\bibinfo {volume} {107}},\ \bibinfo
  {pages} {052403} (\bibinfo {year} {2023})}\BibitemShut {NoStop}%
\bibitem [{\citenamefont {Li}\ \emph {et~al.}(2020)\citenamefont {Li},
  \citenamefont {Qin}, \citenamefont {Zhang}, \citenamefont {Shen},
  \citenamefont {Fei},\ and\ \citenamefont {Fan}}]{Li2020}%
  \BibitemOpen
  \bibfield  {author} {\bibinfo {author} {\bibfnamefont {M.}~\bibnamefont
  {Li}}, \bibinfo {author} {\bibfnamefont {H.}~\bibnamefont {Qin}}, \bibinfo
  {author} {\bibfnamefont {C.}~\bibnamefont {Zhang}}, \bibinfo {author}
  {\bibfnamefont {S.}~\bibnamefont {Shen}}, \bibinfo {author} {\bibfnamefont
  {S.-M.}\ \bibnamefont {Fei}},\ and\ \bibinfo {author} {\bibfnamefont
  {H.}~\bibnamefont {Fan}},\ }\bibfield  {title} {\bibinfo {title}
  {Characterizing multipartite entanglement by violation of chsh
  inequalities},\ }\href {https://doi.org/10.1007/s11128-020-02638-0}
  {\bibfield  {journal} {\bibinfo  {journal} {Quantum Information Processing}\
  }\textbf {\bibinfo {volume} {19}},\ \bibinfo {pages} {142} (\bibinfo {year}
  {2020})}\BibitemShut {NoStop}%
\bibitem [{\citenamefont {Cort{\'e}s-Vega}\ \emph {et~al.}(2023)\citenamefont
  {Cort{\'e}s-Vega}, \citenamefont {Barra}, \citenamefont {Pereira},\ and\
  \citenamefont {Delgado}}]{Cortes-Vega2023}%
  \BibitemOpen
  \bibfield  {author} {\bibinfo {author} {\bibfnamefont {J.}~\bibnamefont
  {Cort{\'e}s-Vega}}, \bibinfo {author} {\bibfnamefont {J.~F.}\ \bibnamefont
  {Barra}}, \bibinfo {author} {\bibfnamefont {L.}~\bibnamefont {Pereira}},\
  and\ \bibinfo {author} {\bibfnamefont {A.}~\bibnamefont {Delgado}},\
  }\bibfield  {title} {\bibinfo {title} {Detecting entanglement of unknown
  states by violating the clauser--horne--shimony--holt inequality},\ }\href
  {https://doi.org/10.1007/s11128-023-03953-y} {\bibfield  {journal} {\bibinfo
  {journal} {Quantum Information Processing}\ }\textbf {\bibinfo {volume}
  {22}},\ \bibinfo {pages} {203} (\bibinfo {year} {2023})}\BibitemShut
  {NoStop}%
\bibitem [{\citenamefont {Werner}(1989)}]{PhysRevA.40.4277}%
  \BibitemOpen
  \bibfield  {author} {\bibinfo {author} {\bibfnamefont {R.~F.}\ \bibnamefont
  {Werner}},\ }\bibfield  {title} {\bibinfo {title} {Quantum states with
  einstein-podolsky-rosen correlations admitting a hidden-variable model},\
  }\href {https://doi.org/10.1103/PhysRevA.40.4277} {\bibfield  {journal}
  {\bibinfo  {journal} {Phys. Rev. A}\ }\textbf {\bibinfo {volume} {40}},\
  \bibinfo {pages} {4277} (\bibinfo {year} {1989})}\BibitemShut {NoStop}%
\bibitem [{\citenamefont {Ure{\~{n}}a}\ \emph {et~al.}(2024)\citenamefont
  {Ure{\~{n}}a}, \citenamefont {Sojo}, \citenamefont {Bermejo-Vega},\ and\
  \citenamefont {Manzano}}]{Urena2024}%
  \BibitemOpen
  \bibfield  {author} {\bibinfo {author} {\bibfnamefont {J.}~\bibnamefont
  {Ure{\~{n}}a}}, \bibinfo {author} {\bibfnamefont {A.}~\bibnamefont {Sojo}},
  \bibinfo {author} {\bibfnamefont {J.}~\bibnamefont {Bermejo-Vega}},\ and\
  \bibinfo {author} {\bibfnamefont {D.}~\bibnamefont {Manzano}},\ }\bibfield
  {title} {\bibinfo {title} {Entanglement detection with classical deep neural
  networks},\ }\href {https://doi.org/10.1038/s41598-024-68213-0} {\bibfield
  {journal} {\bibinfo  {journal} {Scientific Reports}\ }\textbf {\bibinfo
  {volume} {14}},\ \bibinfo {pages} {18109} (\bibinfo {year}
  {2024})}\BibitemShut {NoStop}%
\bibitem [{\citenamefont {Asif}\ \emph {et~al.}(2023)\citenamefont {Asif},
  \citenamefont {Khalid}, \citenamefont {Khan}, \citenamefont {Duong},\ and\
  \citenamefont {Shin}}]{Asif2023}%
  \BibitemOpen
  \bibfield  {author} {\bibinfo {author} {\bibfnamefont {N.}~\bibnamefont
  {Asif}}, \bibinfo {author} {\bibfnamefont {U.}~\bibnamefont {Khalid}},
  \bibinfo {author} {\bibfnamefont {A.}~\bibnamefont {Khan}}, \bibinfo {author}
  {\bibfnamefont {T.~Q.}\ \bibnamefont {Duong}},\ and\ \bibinfo {author}
  {\bibfnamefont {H.}~\bibnamefont {Shin}},\ }\bibfield  {title} {\bibinfo
  {title} {Entanglement detection with artificial neural networks},\ }\href
  {https://doi.org/10.1038/s41598-023-28745-3} {\bibfield  {journal} {\bibinfo
  {journal} {Scientific Reports}\ }\textbf {\bibinfo {volume} {13}},\ \bibinfo
  {pages} {1562} (\bibinfo {year} {2023})}\BibitemShut {NoStop}%
\bibitem [{\citenamefont {Koutný}\ \emph {et~al.}(2023)\citenamefont
  {Koutný}, \citenamefont {Ginés}, \citenamefont {Moczała-Dusanowska},
  \citenamefont {Höfling}, \citenamefont {Schneider}, \citenamefont
  {Predojević},\ and\ \citenamefont {Ježek}}]{doi:10.1126/sciadv.add7131}%
  \BibitemOpen
  \bibfield  {author} {\bibinfo {author} {\bibfnamefont {D.}~\bibnamefont
  {Koutný}}, \bibinfo {author} {\bibfnamefont {L.}~\bibnamefont {Ginés}},
  \bibinfo {author} {\bibfnamefont {M.}~\bibnamefont {Moczała-Dusanowska}},
  \bibinfo {author} {\bibfnamefont {S.}~\bibnamefont {Höfling}}, \bibinfo
  {author} {\bibfnamefont {C.}~\bibnamefont {Schneider}}, \bibinfo {author}
  {\bibfnamefont {A.}~\bibnamefont {Predojević}},\ and\ \bibinfo {author}
  {\bibfnamefont {M.}~\bibnamefont {Ježek}},\ }\bibfield  {title} {\bibinfo
  {title} {Deep learning of quantum entanglement from incomplete
  measurements},\ }\href {https://doi.org/10.1126/sciadv.add7131} {\bibfield
  {journal} {\bibinfo  {journal} {Science Advances}\ }\textbf {\bibinfo
  {volume} {9}},\ \bibinfo {pages} {eadd7131} (\bibinfo {year} {2023})},\
  \Eprint
  {https://arxiv.org/abs/https://www.science.org/doi/pdf/10.1126/sciadv.add7131}
  {https://www.science.org/doi/pdf/10.1126/sciadv.add7131} \BibitemShut
  {NoStop}%
\bibitem [{\citenamefont {Qu}\ \emph {et~al.}(2023)\citenamefont {Qu},
  \citenamefont {Zhang}, \citenamefont {Shen}, \citenamefont {Yu},\ and\
  \citenamefont {Li}}]{Qu2023}%
  \BibitemOpen
  \bibfield  {author} {\bibinfo {author} {\bibfnamefont {Y.-D.}\ \bibnamefont
  {Qu}}, \bibinfo {author} {\bibfnamefont {R.-Q.}\ \bibnamefont {Zhang}},
  \bibinfo {author} {\bibfnamefont {S.-Q.}\ \bibnamefont {Shen}}, \bibinfo
  {author} {\bibfnamefont {J.}~\bibnamefont {Yu}},\ and\ \bibinfo {author}
  {\bibfnamefont {M.}~\bibnamefont {Li}},\ }\bibfield  {title} {\bibinfo
  {title} {Entanglement detection with complex-valued neural networks},\ }\href
  {https://doi.org/10.1007/s10773-023-05460-3} {\bibfield  {journal} {\bibinfo
  {journal} {International Journal of Theoretical Physics}\ }\textbf {\bibinfo
  {volume} {62}},\ \bibinfo {pages} {206} (\bibinfo {year} {2023})}\BibitemShut
  {NoStop}%
\bibitem [{\citenamefont {Luo}\ \emph {et~al.}(2023)\citenamefont {Luo},
  \citenamefont {Liu},\ and\ \citenamefont {Zhang}}]{PhysRevA.108.052424}%
  \BibitemOpen
  \bibfield  {author} {\bibinfo {author} {\bibfnamefont {Y.-J.}\ \bibnamefont
  {Luo}}, \bibinfo {author} {\bibfnamefont {J.-M.}\ \bibnamefont {Liu}},\ and\
  \bibinfo {author} {\bibfnamefont {C.}~\bibnamefont {Zhang}},\ }\bibfield
  {title} {\bibinfo {title} {Detecting genuine multipartite entanglement via
  machine learning},\ }\href {https://doi.org/10.1103/PhysRevA.108.052424}
  {\bibfield  {journal} {\bibinfo  {journal} {Phys. Rev. A}\ }\textbf {\bibinfo
  {volume} {108}},\ \bibinfo {pages} {052424} (\bibinfo {year}
  {2023})}\BibitemShut {NoStop}%
\bibitem [{\citenamefont {Roik}\ \emph {et~al.}(2022)\citenamefont {Roik},
  \citenamefont {Bartkiewicz}, \citenamefont {Černoch},\ and\ \citenamefont
  {Lemr}}]{ROIK2022128270}%
  \BibitemOpen
  \bibfield  {author} {\bibinfo {author} {\bibfnamefont {J.}~\bibnamefont
  {Roik}}, \bibinfo {author} {\bibfnamefont {K.}~\bibnamefont {Bartkiewicz}},
  \bibinfo {author} {\bibfnamefont {A.}~\bibnamefont {Černoch}},\ and\
  \bibinfo {author} {\bibfnamefont {K.}~\bibnamefont {Lemr}},\ }\bibfield
  {title} {\bibinfo {title} {Entanglement quantification from collective
  measurements processed by machine learning},\ }\href
  {https://doi.org/https://doi.org/10.1016/j.physleta.2022.128270} {\bibfield
  {journal} {\bibinfo  {journal} {Physics Letters A}\ }\textbf {\bibinfo
  {volume} {446}},\ \bibinfo {pages} {128270} (\bibinfo {year}
  {2022})}\BibitemShut {NoStop}%
\bibitem [{\citenamefont {Ma}\ and\ \citenamefont {Yung}(2018)}]{Ma2018}%
  \BibitemOpen
  \bibfield  {author} {\bibinfo {author} {\bibfnamefont {Y.-C.}\ \bibnamefont
  {Ma}}\ and\ \bibinfo {author} {\bibfnamefont {M.-H.}\ \bibnamefont {Yung}},\
  }\bibfield  {title} {\bibinfo {title} {Transforming bell's inequalities into
  state classifiers with machine learning},\ }\href
  {https://doi.org/10.1038/s41534-018-0081-3} {\bibfield  {journal} {\bibinfo
  {journal} {npj Quantum Information}\ }\textbf {\bibinfo {volume} {4}},\
  \bibinfo {pages} {34} (\bibinfo {year} {2018})}\BibitemShut {NoStop}%
\bibitem [{\citenamefont {Greenwood}\ \emph {et~al.}(2023)\citenamefont
  {Greenwood}, \citenamefont {Wu}, \citenamefont {Zhu}, \citenamefont {Kirby},\
  and\ \citenamefont {Qian}}]{PhysRevApplied.19.034058}%
  \BibitemOpen
  \bibfield  {author} {\bibinfo {author} {\bibfnamefont {A.~C.}\ \bibnamefont
  {Greenwood}}, \bibinfo {author} {\bibfnamefont {L.~T.}\ \bibnamefont {Wu}},
  \bibinfo {author} {\bibfnamefont {E.~Y.}\ \bibnamefont {Zhu}}, \bibinfo
  {author} {\bibfnamefont {B.~T.}\ \bibnamefont {Kirby}},\ and\ \bibinfo
  {author} {\bibfnamefont {L.}~\bibnamefont {Qian}},\ }\bibfield  {title}
  {\bibinfo {title} {Machine-learning-derived entanglement witnesses},\ }\href
  {https://doi.org/10.1103/PhysRevApplied.19.034058} {\bibfield  {journal}
  {\bibinfo  {journal} {Phys. Rev. Appl.}\ }\textbf {\bibinfo {volume} {19}},\
  \bibinfo {pages} {034058} (\bibinfo {year} {2023})}\BibitemShut {NoStop}%
\bibitem [{\citenamefont {Kookani}\ \emph {et~al.}(2024)\citenamefont
  {Kookani}, \citenamefont {Mafi}, \citenamefont {Kazemikhah}, \citenamefont
  {Aghababa}, \citenamefont {Fouladi},\ and\ \citenamefont
  {Barati}}]{Kookani2024}%
  \BibitemOpen
  \bibfield  {author} {\bibinfo {author} {\bibfnamefont {A.}~\bibnamefont
  {Kookani}}, \bibinfo {author} {\bibfnamefont {Y.}~\bibnamefont {Mafi}},
  \bibinfo {author} {\bibfnamefont {P.}~\bibnamefont {Kazemikhah}}, \bibinfo
  {author} {\bibfnamefont {H.}~\bibnamefont {Aghababa}}, \bibinfo {author}
  {\bibfnamefont {K.}~\bibnamefont {Fouladi}},\ and\ \bibinfo {author}
  {\bibfnamefont {M.}~\bibnamefont {Barati}},\ }\bibfield  {title} {\bibinfo
  {title} {Xpookynet: advancement in quantum system analysis through
  convolutional neural networks for detection of entanglement},\ }\href
  {https://doi.org/10.1007/s42484-024-00183-y} {\bibfield  {journal} {\bibinfo
  {journal} {Quantum Machine Intelligence}\ }\textbf {\bibinfo {volume} {6}},\
  \bibinfo {pages} {50} (\bibinfo {year} {2024})}\BibitemShut {NoStop}%
\bibitem [{\citenamefont {Hensen}\ \emph {et~al.}(2015)\citenamefont {Hensen},
  \citenamefont {Bernien}, \citenamefont {Dr{\'e}au}, \citenamefont {Reiserer},
  \citenamefont {Kalb}, \citenamefont {Blok}, \citenamefont {Ruitenberg},
  \citenamefont {Vermeulen}, \citenamefont {Schouten}, \citenamefont
  {Abell{\'a}n}, \citenamefont {Amaya}, \citenamefont {Pruneri}, \citenamefont
  {Mitchell}, \citenamefont {Markham}, \citenamefont {Twitchen}, \citenamefont
  {Elkouss}, \citenamefont {Wehner}, \citenamefont {Taminiau},\ and\
  \citenamefont {Hanson}}]{Hensen2015}%
  \BibitemOpen
  \bibfield  {author} {\bibinfo {author} {\bibfnamefont {B.}~\bibnamefont
  {Hensen}}, \bibinfo {author} {\bibfnamefont {H.}~\bibnamefont {Bernien}},
  \bibinfo {author} {\bibfnamefont {A.~E.}\ \bibnamefont {Dr{\'e}au}}, \bibinfo
  {author} {\bibfnamefont {A.}~\bibnamefont {Reiserer}}, \bibinfo {author}
  {\bibfnamefont {N.}~\bibnamefont {Kalb}}, \bibinfo {author} {\bibfnamefont
  {M.~S.}\ \bibnamefont {Blok}}, \bibinfo {author} {\bibfnamefont
  {J.}~\bibnamefont {Ruitenberg}}, \bibinfo {author} {\bibfnamefont {R.~F.~L.}\
  \bibnamefont {Vermeulen}}, \bibinfo {author} {\bibfnamefont {R.~N.}\
  \bibnamefont {Schouten}}, \bibinfo {author} {\bibfnamefont {C.}~\bibnamefont
  {Abell{\'a}n}}, \bibinfo {author} {\bibfnamefont {W.}~\bibnamefont {Amaya}},
  \bibinfo {author} {\bibfnamefont {V.}~\bibnamefont {Pruneri}}, \bibinfo
  {author} {\bibfnamefont {M.~W.}\ \bibnamefont {Mitchell}}, \bibinfo {author}
  {\bibfnamefont {M.}~\bibnamefont {Markham}}, \bibinfo {author} {\bibfnamefont
  {D.~J.}\ \bibnamefont {Twitchen}}, \bibinfo {author} {\bibfnamefont
  {D.}~\bibnamefont {Elkouss}}, \bibinfo {author} {\bibfnamefont
  {S.}~\bibnamefont {Wehner}}, \bibinfo {author} {\bibfnamefont {T.~H.}\
  \bibnamefont {Taminiau}},\ and\ \bibinfo {author} {\bibfnamefont
  {R.}~\bibnamefont {Hanson}},\ }\bibfield  {title} {\bibinfo {title}
  {Loophole-free bell inequality violation using electron spins separated by
  1.3 kilometres},\ }\href {https://doi.org/10.1038/nature15759} {\bibfield
  {journal} {\bibinfo  {journal} {Nature}\ }\textbf {\bibinfo {volume} {526}},\
  \bibinfo {pages} {682} (\bibinfo {year} {2015})}\BibitemShut {NoStop}%
\bibitem [{\citenamefont {Cirel'son}(1980)}]{Cirelson1980}%
  \BibitemOpen
  \bibfield  {author} {\bibinfo {author} {\bibfnamefont {B.~S.}\ \bibnamefont
  {Cirel'son}},\ }\bibfield  {title} {\bibinfo {title} {Quantum generalizations
  of bell's inequality},\ }\href {https://doi.org/10.1007/BF00417500}
  {\bibfield  {journal} {\bibinfo  {journal} {Letters in Mathematical Physics}\
  }\textbf {\bibinfo {volume} {4}},\ \bibinfo {pages} {93} (\bibinfo {year}
  {1980})}\BibitemShut {NoStop}%
\bibitem [{\citenamefont {Chruściński}\ and\ \citenamefont
  {Sarbicki}(2014)}]{Chr_2014}%
  \BibitemOpen
  \bibfield  {author} {\bibinfo {author} {\bibfnamefont {D.}~\bibnamefont
  {Chruściński}}\ and\ \bibinfo {author} {\bibfnamefont {G.}~\bibnamefont
  {Sarbicki}},\ }\bibfield  {title} {\bibinfo {title} {Entanglement witnesses:
  construction, analysis and classification},\ }\href
  {https://doi.org/10.1088/1751-8113/47/48/483001} {\bibfield  {journal}
  {\bibinfo  {journal} {Journal of Physics A: Mathematical and Theoretical}\
  }\textbf {\bibinfo {volume} {47}},\ \bibinfo {pages} {483001} (\bibinfo
  {year} {2014})}\BibitemShut {NoStop}%
\bibitem [{\citenamefont {Simon}(2000)}]{PhysRevLett.84.2726}%
  \BibitemOpen
  \bibfield  {author} {\bibinfo {author} {\bibfnamefont {R.}~\bibnamefont
  {Simon}},\ }\bibfield  {title} {\bibinfo {title} {Peres-horodecki
  separability criterion for continuous variable systems},\ }\href
  {https://doi.org/10.1103/PhysRevLett.84.2726} {\bibfield  {journal} {\bibinfo
   {journal} {Phys. Rev. Lett.}\ }\textbf {\bibinfo {volume} {84}},\ \bibinfo
  {pages} {2726} (\bibinfo {year} {2000})}\BibitemShut {NoStop}%
\bibitem [{\citenamefont {T\'oth}(2005)}]{PhysRevA.71.010301}%
  \BibitemOpen
  \bibfield  {author} {\bibinfo {author} {\bibfnamefont {G.}~\bibnamefont
  {T\'oth}},\ }\bibfield  {title} {\bibinfo {title} {Entanglement witnesses in
  spin models},\ }\href {https://doi.org/10.1103/PhysRevA.71.010301} {\bibfield
   {journal} {\bibinfo  {journal} {Phys. Rev. A}\ }\textbf {\bibinfo {volume}
  {71}},\ \bibinfo {pages} {010301} (\bibinfo {year} {2005})}\BibitemShut
  {NoStop}%
\bibitem [{\citenamefont {Dowling}\ \emph {et~al.}(2004)\citenamefont
  {Dowling}, \citenamefont {Doherty},\ and\ \citenamefont
  {Bartlett}}]{PhysRevA.70.062113}%
  \BibitemOpen
  \bibfield  {author} {\bibinfo {author} {\bibfnamefont {M.~R.}\ \bibnamefont
  {Dowling}}, \bibinfo {author} {\bibfnamefont {A.~C.}\ \bibnamefont
  {Doherty}},\ and\ \bibinfo {author} {\bibfnamefont {S.~D.}\ \bibnamefont
  {Bartlett}},\ }\bibfield  {title} {\bibinfo {title} {Energy as an
  entanglement witness for quantum many-body systems},\ }\href
  {https://doi.org/10.1103/PhysRevA.70.062113} {\bibfield  {journal} {\bibinfo
  {journal} {Phys. Rev. A}\ }\textbf {\bibinfo {volume} {70}},\ \bibinfo
  {pages} {062113} (\bibinfo {year} {2004})}\BibitemShut {NoStop}%
\bibitem [{\citenamefont {Ho}\ and\ \citenamefont
  {Imoto}(2017)}]{PhysRevA.95.032135}%
  \BibitemOpen
  \bibfield  {author} {\bibinfo {author} {\bibfnamefont {L.~B.}\ \bibnamefont
  {Ho}}\ and\ \bibinfo {author} {\bibfnamefont {N.}~\bibnamefont {Imoto}},\
  }\bibfield  {title} {\bibinfo {title} {Generalized modular-value-based scheme
  and its generalized modular value},\ }\href
  {https://doi.org/10.1103/PhysRevA.95.032135} {\bibfield  {journal} {\bibinfo
  {journal} {Phys. Rev. A}\ }\textbf {\bibinfo {volume} {95}},\ \bibinfo
  {pages} {032135} (\bibinfo {year} {2017})}\BibitemShut {NoStop}%
\bibitem [{\citenamefont {Collins}\ \emph {et~al.}(2002)\citenamefont
  {Collins}, \citenamefont {Gisin}, \citenamefont {Linden}, \citenamefont
  {Massar},\ and\ \citenamefont {Popescu}}]{PhysRevLett.88.040404}%
  \BibitemOpen
  \bibfield  {author} {\bibinfo {author} {\bibfnamefont {D.}~\bibnamefont
  {Collins}}, \bibinfo {author} {\bibfnamefont {N.}~\bibnamefont {Gisin}},
  \bibinfo {author} {\bibfnamefont {N.}~\bibnamefont {Linden}}, \bibinfo
  {author} {\bibfnamefont {S.}~\bibnamefont {Massar}},\ and\ \bibinfo {author}
  {\bibfnamefont {S.}~\bibnamefont {Popescu}},\ }\bibfield  {title} {\bibinfo
  {title} {Bell inequalities for arbitrarily high-dimensional systems},\ }\href
  {https://doi.org/10.1103/PhysRevLett.88.040404} {\bibfield  {journal}
  {\bibinfo  {journal} {Phys. Rev. Lett.}\ }\textbf {\bibinfo {volume} {88}},\
  \bibinfo {pages} {040404} (\bibinfo {year} {2002})}\BibitemShut {NoStop}%
\end{thebibliography}%

%\end{widetext}
\end{document}